# Past, Present and Future of Hadoop: A Survey

Ameneh Zarei[b],*, Shahla Safari[a], Mahmood Ahmadi[b], Farhad Mardukhi[b]

[a] *Computer Engineering Department, Zagros University of Kermanshah, Iran*
[b] *Computer Engineering Department, Razi University, Kermanshah, Iran*

Abstract

In this paper, a technology for massive data storage and computing named Hadoop is surveyed. Hadoop consists of heterogeneous computing devices like regular PCs abstracting away the details of parallel processing and developers can just concentrate on their computational problem. A Hadoop cluster is made of two parts: HDFS and MapReduce. Hadoop cluster uses HDFS for data management. HDFS provides storage for input and output data in MapReduce jobs and is designed with abilities like high – fault tolerance, high – distribution capacity and high – throughput. It is also suitable for storing Terabyte or Petabyte data on cluster and it runs on flexible hardware like commodity devices.

1. Introduction

Nowadays, the organizations tend to data-intensive distributed applications dealing with processing and transmitting their massive data, namely Big Data, through an efficient way. These data should be handled efficiently and quickly, so there are a number of technologies to deal with "Big Data" including storing managing, transmitting and processing of data. For example, Google provides three software techniques for massive multimedia data storage: Google File System (GFS), MapReduce and BigTable. Another known technology is Apache Hadoop. It is an open source framework which implemented GFS and MapReduce. Hadoop is a distributed file system for storing and executing distributed applications [1], [2]. Hadoop becomes most popular system for processing large data sets [3], [4], [5] where nodes simultaneously perform computing and storing functions [6]. Hadoop is especially suitable for high - throughput and write – once – read – many applications. It is adopted by Internet Companies like Yahoo, Taobao and Baidu to support their clusters which encompasses thousands of nodes for log analysing, data mining and searching and especially for handling Big Data. Hadoop demonstrates in both scientific computing and mass data processing. The performance of Hadoop programs depends on hardware and software configurations.

Apache Hadoop has become massive data management open source software which is adopted for situations we need to deal with very big amount of data.

* Tel: +989189281350
Email: a.zarei@zagros-ks.ac.ir (Ameneh Zarei)
s.safari@zagros-ks.ac.ir (Shahla Safari)
m.ahmadi@razi.ac.ir (Mahmood Ahmadi)
mardukhi@razi.ac.ir (Farhad Mardukhi)

But in some scenarios Hadoop is particularly useful:

- To process complex information ( through parallel algorithms )
- To process non – structural data
- To getting an unessential result
- To deal with machine learning tasks
- To fit large Datasets in RAM or hard disk
- To process data within too many cores

Hadoop is an implementation of reference MapReduce for commodity clusters and enterprise data centers. In addition it has been extended to other fields, such as Hyrax for Android smartphone applications where data and computations are distributed [7].

Hadoop has two main parts: Hadoop distributed file system (HDFS) and Hadoop MapReduce framework.

HDFS is a distributed file system with master-slave architecture. It was designed by GFS (Google file system) which provides high – throughput access to application data. A daemon program named "NameNode" runs on the master node, manages system Metadata and logically divides files into equal sized blocks and controls their distribution on cluster. It does this by file replication factor for fault tolerance. Several "DataNode" daemons running on slave nodes store actual data blocks and execute management tasks assigned to them by "NameNode" and deal with read – write requests form users.

Hadoop MapReduce framework runs on top of HDFS and depends on the conventional master – slave architecture. The Master node executes a single daemon named JobTracker to manage jobs status and assign tasks. On the slave node, a TaskTracker daemon is responsible for launching new JVM processes to perform a task while reporting task processes and idle task slots to JobTracker, through heartbeat signals. (Slot number is the maximum number of MapReduce tasks which can run on a slave node concurrently.) Then JobTracker updates TaskTracker states and assigns new tasks to it due to available slots and data locality.

Some data management services use Hadoop as baseline technology to exploit its advanced designed aspects. Some services use hadoop as back – end processing system and extend it to enrich their purposes [8]. Hadoop has been used for solving many problems and in areas that need analysing and processing Big Data such as data screening for phylogenetic analysis and etc [9], [10].

Our motivation to write this survey is that we found MapReduce most important distributed processing model for processing massive data. Hadoop is an open – source implementation of MapReduce which is widely used. As we tried to find a survey that inspects it we came to this result that there is no complete survey that covers most of basic ideas in designing Hadoop and also introduce its challenges and describe new trends and optimizations in it.

Scope of the survey

In the Section 2 we introduce MapReduce, Section 3 presents Hadoop ecosystem, in Section 4 we review the HDFS, in Section 5 some of HDFS failures and optimizations are considered, in Section 6 we introduce Hadoop schedulers, Section 7 illustrates Hadoop implementation on Cloud, Grid and GPU, Section 8 inspects security in Hadoop, in Section 9 we present some of Hadoop platforms and usages in science and technology, Section 10 is about Hadoop problems and challenges and Section 11 is the conclusion of this paper.

2. What is MapReduce?

Currently, the existence of applications which process massive data impose us to leverage parallel programs regarding too limited speed of a single processor. Experiments have implied that the most time consuming parts are encompassing those operations which are independent and can be processed collectively (reductively). This issue illustrates the importance of MapReduce paradigm [11], [12], [13], [14], [15] which provides the capability of transforming a typical sequential code into corresponding parallel code automatically [16], [17].

MapReduce [18] is for Big Data sets that need to be indexed, categorized, sorted, culled, analysed, etc. Looking through each record in a serial environment can take very long time. MapReduce lets data to distribute on a large cluster and can distribute tasks across the data sets to work on pieces of data independently and simultaneously. This allows Big Data to be processed in a short time. Data – intensive operations require so many I/O operations and massive amount of computing. In addition of CPU and storage systems, network system has an important effect on the performance of MapReduce system, so efficiently exploiting these resources require right architecture [19], [20], [21].

Google implemented MapReduce platform and uses it propriety and unleashed its full power [5], [17], [22], [23]. Apache has produced an open source MapReduce platform named Hadoop. Google's MapReduce makes the complex process of parallel execution of an application over a lot of servers and nodes transparent to programmers [24], [25]. In addition of dealing with Big Data, MapReduce is used in areas that need parallel processing; for example, it is used as parallel computing framework for smart phones and mobiles [26].

Figure 1 depicts how Hadoop deals with Big Data sets. MapReduce model divides parallel computing process into two functions: Map function and Reduce function. Map function takes a pair of (key, value) as input and returns one or more intermediate state (key, value) pair sets. When a job is submitted to MapReduce framework, MapReduce splits it into several map tasks and submits them to different nodes for running. Each map task deals with only one portion of input. Finishing map task running the results will be sent to reduce function in terms of a set of intermediate (key, value) pairs. Reduce function integrates pairs based on a specific key and then generates and returns (key, value) pairs that user client requires [2], [12] [18],[27],[28]. There is no parallelism between Map tasks and Reduce tasks because Map tasks are followed by Reduce tasks [29]. Each MapReduce step is a job in Hadoop and every program included several jobs running as a pipeline [30].

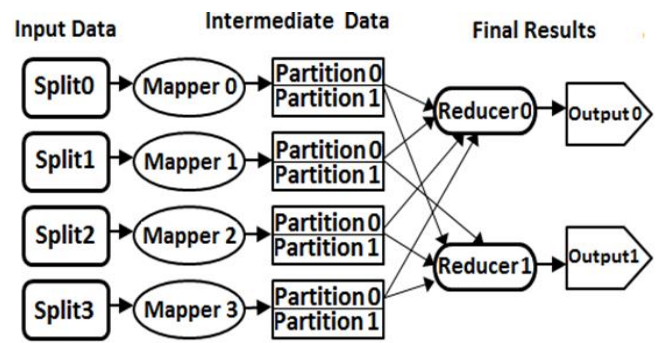

Fig 1: Execution of MapReduce.

MapReduce programming model borrows two concepts *map* and *reduce* from the other languages (e.g. Lisp) to handle list processing. This is for separating computing expressions of user's applications from details of huge parallel data processing. It is realized through the following three phases: The input reading phase to inputs (usually from a distributed file system) and parses it into (key, value) pair records. Map function replicates these records and maps each of them to a set of intermediate (key, value) pairs. In the second phase, the intermediate pairs are partitioned by partitioning function, then sorted and grouped according to their keys. There can be an optional function to reduce the size of intermediate data. In the third phase the reduce function, reduces the results of the previous phase for each unique key in a sorted list, to achieve a final result [31].

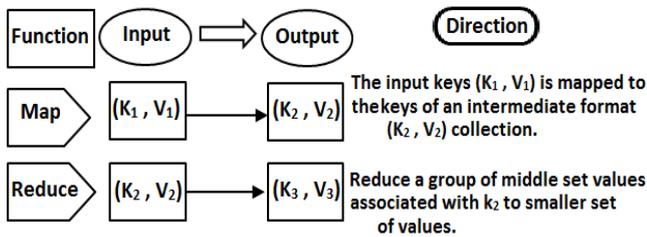

Fig 2: Processing Distributed Data in MapReduce.

Figure 2, depicts the processing of distributed data in MapReduce. There might be several stages of intermediate processing where the results of each stage would be the input for the next stage.

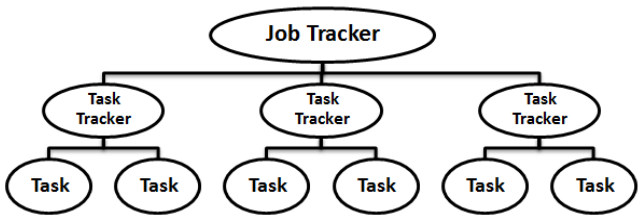

Fig 3: Relationship of Tasks Allocation in MapReduce.

MapReduce function is scheduled by JobTracker and TaskTracker. The relationship of allocating tasks is shown in Figure 3. JobTracker which is in NameNode, is the only master control which can run on any computer in the cluster. It schedules and manages other TaskTrackers, monitors the condition of tasks, allocates the map and reduces tasks to free TaskTrackers for running in parallel. There can be several TaskTrackers. TaskTracker is responsible for implementing the tasks which is running on DataNode. It means that DataNode is not only a storage node, but also is a computing node. If a TaskTracker's task fails, JobTracker assigns that task to another free TaskTracker. Hadoop tries to shorten the execution time cost of MapReduce jobs [32].

3. Hadoop Ecosystem

Hadoop is a collection of several projects which are mostly introduced in this survey, as shown some of them in Figure 4 and overviewed below.

Hue is an open – source web interface that supports Apache Hadoop and its ecosystem. Hue has a file browser for HDFS – a job browser for MapReduce/YARN – Hbase browser, Zookeeper browser, and query editors for Hive, Pig, Cloudera Impala and Sqoop2. It also works with Oozie applications for creating and monitoring workflows.

*Beeswax* application enables user to imply and run queries on Apache Hive which is data warehousing system.

*Sqoop* is a command – line interface application which efficiently transfers BigData between relational databases and Hadoop. It can be used for importing data from RDBMSs to HDFS and vice versa [33].

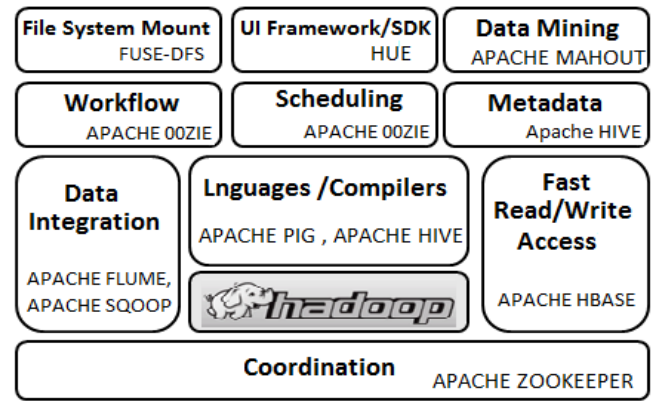

Fig 4: Hadoop Ecosystem.

*Flume* is a framework which provides some reliable services to gathering and transferring BigData [34].

*Apache Mahout* is one of Apache projects which its algorithms are implemented on top of Hadoop. It uses MapReduce programming model to provide scalable machine learning algorithms. Mahout can be used for clustering, classification, association rule mining, batch based collaborative filtering and etc [35].

*Fuse* is a user – level file system that consists of a kernel module and a userland daemon. Applications and userland daemons negotiate through kernel module to forward requests for file system access to the daemon and returning the results of the daemon to the application [36].

Other parts of this ecosystem will be discussed in next sections.

4. Hadoop Distributed File System (HDFS)

HDFS is an Apache software project and is a sub – project of another project named Hadoop. Hadoop is ideal for huge data volume (e.g. Petabytes and Terabytes) which uses HDFS as its storage system. HDFS has been successfully implemented for clusters with 10 – 4500 nodes and is able to store up to 25 Petabytes of data. It allows users to connect to nodes (commodity personal computers) contained within clusters which data files are distributed on them. Then users can have access to the data files as a seamless file system and store their data. Data files access is in a streaming manner, which means applications or commands are executed directly using MapReduce model [5], [12], [21], [37].

Hadoop distributed file system (HDFS) is designed to run on commodity hardware. It has many similarities to other distributed file systems but still has many differences as well. HDFS has high – fault tolerance and is designed to be deployed on low – cost hardware.

HDFS provides high – throughput access to application data and is properly beneficial for Big Data sets. HDFS is designed to store very large data sets reliably and to stream those data sets on high bandwidth to user's applications. In a large cluster, thousands of servers are attached together to storing data and executing user's application tasks. By distributing storage and computation across many servers, the resources can grow with demand while remaining economical on every size. The main purpose of HDFS is storing data reliably in the presence of failures.

Hadoop becomes the most popular platform for content classifications in World Wide Web which is an easy use software platform to develop and handle large – scale data. Experiments have shown that when amount of data for dealing with a massive class file, is about 18 Gigabyte, the stand – alone consumed time is 4.8 times for the Hadoop cluster. Hadoop accesses data in the form of stream, providing a very high transfer rate which dedicates efficient and even perfect performance [27].

A Hadoop job includes parallel Map tasks, a shuffle and sort phase and parallel reduce tasks which execute after shuffling phase completes.

The users who use MapReduce tasks and Hadoop framework should write codes to manage sorting, shuffling and coordination of tasks run in a parallel manner [38]. For the sake of fault tolerance, Hadoop has multiple trials of execution for each task which are called Attempts. When one Attempt fails, another one of the same task automatically starts. This will restart the process from the beginning. The whole restarting process continuous until the task is completed or the number of failed Attempts exceeds a threshold. However, the speculative jobs are exceptions. The word speculative job is used when the performance of Hadoop may be reduced due to slow running of some tasks or for handling strangler tasks. In such cases, Hadoop replicates these special tasks on idle nodes. When one of the replicated tasks or the original tasks is finished, Hadoop commits its results and kills the other replicas. The decision of killing is made during run time and it's not one of the deployment failures [30],[39].

In HDFS it is often better to move the computation to where the data is stored rather than move the data to where the application is running regarding to huge volume of data. This will minimize network traffic and increases the overall throughput of the system. Figure 5 shows a simplified overlook of HDFS structure.

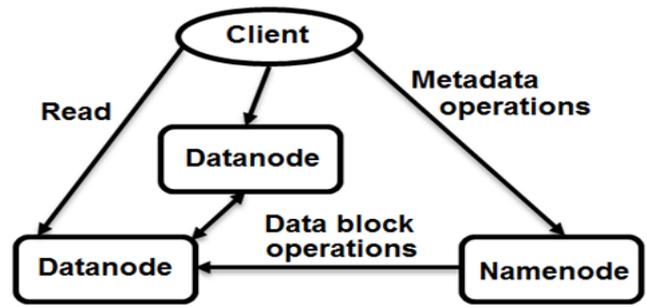

Fig 5: The HDFS File Structure.

HDFS provides interfaces for applications to move themselves closer to where data is stored. This will be done by data awareness feature that we explain later [23]. In other world, Hadoop moves codes to where data are stored. It makes Hadoop more scalable [31].

HDFS is designed to be portable from one platform to another easily. This makes HDFS an appropriate platform to choose for Big Data sets for applications.

HDFS has many similarities to other distributed file systems, but it is different in some aspects. An important difference is that HDFS is a write - once – read – many model which relaxes concurrency control requirements and simplifies data coherency and enables high - throughput access.

HDFS limits writing of data to one writer per time rigorously. Bytes are always appended to the end of the stream and guaranteed to be stored in order of the writing.

HDFS has many goals; here are some of most notable of them:

- Fault tolerance : by detecting faults and doing a quick automatic recovery
- Data access through MapReduce stream
- Simple and powerful coherency model
- Processing logic close to data instead of data close to processing logic
- Portability on heterogeneous hardware and operating systems
- Scalability for reliable storage and processing large amount of data
- Economy by data distributing and processing data on clusters of commodity computers
- Efficiency by distributing data and logic to process it in parallel on nodes where data is stored
- Reliability by automatically maintaining several copies of data and automatically redeploying processing in failure events.

HDFS provides interfaces for applications to make them closer to where data is located. HDFS can be accessed in many different ways. It provides native Java application programming interface (API) and a native C – language wrapper for Java API. In addition user can use a web browser to brows HDFS files.

As shown in Figure 6 user application accesses the file system by HDFS client which is a code library that exports into the HDFS file system interface [40].

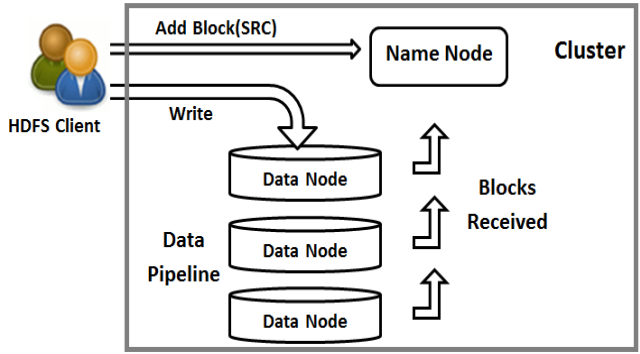

Fig 6: User Interface in Hadoop.

In order to reading a HDFS file, user applications use a standard Java file input stream as if the data is in the native file system. To writing data on HDFS, user applications use a standard Java file output stream. The data stream is first fragmented into equal – sized 64 Megabyte data block and then the smaller packets by the client thread.

*4.1 HDFS architecture*

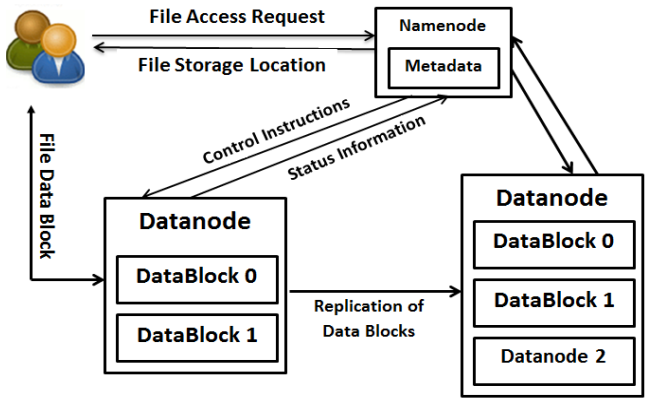

Fig 7: HDFS Architecture.

Figure 7 shows HDFS architecture [41]. HDFS provides file name space for users. One file splits into several data blocks and stores on sets on DataNodes. User will learn the DataNode location where data bases are stored on it through NameNode and then read and write data directly from DataNode. NameNode doesn't involve in file transitions but is responsible of uniform dispatcher of DataNodes to duplicate, delete and create data blocks. HDFS has a master – slave structure. Each HDFS cluster has a master node named NameNode. A master server manages file name space and regulates client access to files. A NameNode is arbitrator and repository for all HDFS Metadata. In addition there is several DataNode in this structure. Topically for each node in a cluster, there is DataNode managing storage attached to the nodes that they run on. HDFS exposes a file system name space and allows user data to be stored in files in name space. A file divides into several data blocks which are stored in DataNodes. NameNode executes file system name space operations like opening, renaming and closing files and directories. Also NameNode maps data blocks to DataNodes. DataNodes are responsible for client's read and write requests. They also create, delete and replicate data blocks upon instructions from NameNode. The Figure 8 illustrates HDFS architecture with more clarifications:[42]

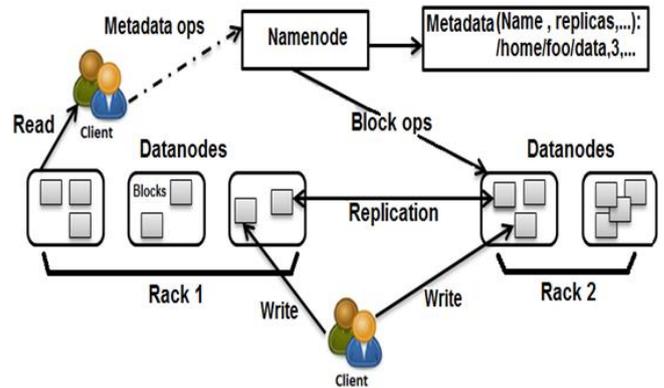

Fig 8: HDFS Architecture.

In HDFS structure, each file divides into several 64 Megabyte sized blocks and stores in DataNodes. NameNode keeps Metadata and provides query service for clients [43]. Figure 9 represents a high level lookup of HDFS structure.

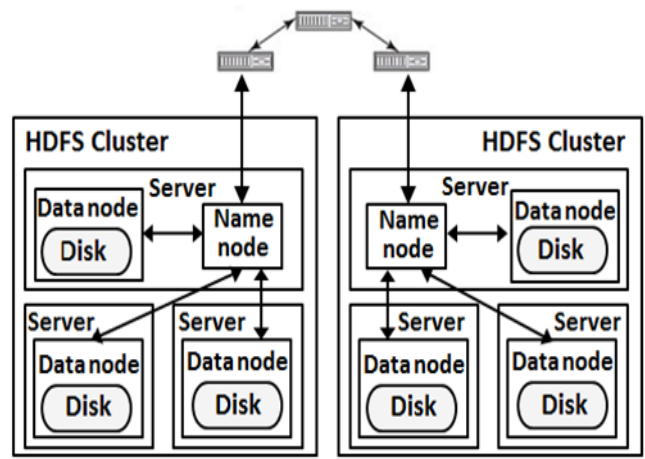

Fig 9: HDFS Structure.

All HDFS communication protocols are based on TCP/IP. HDFS clients connect to a transmission control protocol (TCP) port which is opened on the NameNode and then communicate with the NameNode using remote procedure call (RPC) – based protocol (unlike other interfaces like MPI) DataNodes talk to NameNode by propriety block based protocol.

DataNodes can communicate to each other to move copies of data and information, to rebalance data distribution or to make sure that replication of data is high enough [23].

HDFS supports traditional hierarchical file organization which a user or an application can create directories and store files inside them. The file name space hierarchy is similar to most other existing file systems. Users can create files, move or rename them or delete them. HDFS also supports third – party file systems like Cloud store or Amazon simple storage service (s3).

*4.2 Data replication*

HDFS is built based on Java language. Any machine which supports Java can execute DataNode and name space software. Using high – portable Java language means that HDFS can be deployed on a wide range of machines. NameNode keeps the file system name space. Any changes in file system name space or its properties are recorded by the NameNode. As we mentioned before, data in HDFS are divided to be processed in parallel, each portion of divided data will be copied at least 3 times and will be distributed across the cluster. Each copy of data is named replica. An application can specify number of file replicas that should be maintained by the HDFS. The number of replicas of a file is named replication factor. This information is stored in the NameNode [44]. In general deployment of Hadoop, the NameNode backs up three replicas of blocks and assigns each of them to a different DataNode [1].

HDFS is designed for storing a huge amount of data across machines in a large cluster and stores each file as a sequence of blocks. Every block has an equal size except the last block. The blocks of a file are replicated for fault tolerance. Block size and replication factor for each block is configurable. An application can specify the number of replicas of a file. Replication factor can be specified or configured at the file creation time and can be changed later. The files on HDFS are write – once and have strictly one writer at a time. By default, two copies of each block are stored by different DataNodes on a same rack and a third copy is stored on a DataNode in another rack (for the sake of reliability) [23].

NameNode makes all decisions about the replications of blocks. It periodically receives a heartbeat and a block report from each DataNode in the cluster. Arriving of a heartbeat from DataNode means that DataNode is working properly, in other words, it means that node is alive. A block report is consisting of a list from all blocks and a DataNode. The need for re – replication might increase for several reasons: a DataNode might become unavailable, a replica might become corrupted, a hard disk on a DataNode might crash or replication factor for a data block might be increased.

The placement of replicas is very important for HDFS reliability and performance. The optimization of replacement of replicas, distinguished HDFS from other file systems. This feature needs a lot of tuning and experience. The purpose of having a rack – aware replica policy is to improve reliability, network bandwidth utilization and availability. The existing implementation for the replica placement is the first step for this direction. The short term goals for this policy is to validate it on production systems and learn more about their behaviour to build a foundation to test and research more sophisticated policy.

In general the first replica is located in a random node; the second replica is placed in a random node in a different rack and the third replica is placed in a different node in the same rack of the second replica. Other replicas will be placed on random nodes across the cluster [45].

When a client creates a file in HDFS, it first caches the data into a temporary local file. It then redirects subsequent writes to the temporary file. When the temporary file accumulates enough data to fill one HDFS block, the client reports this to the NameNode which converts the file to a permanent DataNode. The client then closes the temporary file and flushes any remaining data to the newly created DataNode. The NameNode then commits the DataNode to disk.

On Startup, NameNode is in a state named safe mode. Replication or deletion of the data blocks does not occur in the safe mood state and the file system is read – only. Also an administrator can make a NameNode to enter the safe mode manually. NameNode in safe mode receives heartbeats and block reports from DataNodes. Block report consists of a list of data blocks that a data block is hosting. Each block has a specified minimum number of replicas. When the number of replicated data blocks has checked in NameNode, a block is considered safely replicated. After a configurable percentage of safely replicated data blocks checks in with the NameNode, (plus an additional time like 30 seconds), NameNode exits the safe mode state. Then NameNode verifies the list of data blocks that their number of replicas is less than a threshold and then NameNode replicates these blocks to other DataNodes [46], [47].

In the safe mode the image file will be loaded into memory and will replace the edit log and by this the state of the file system will be reconstructed. Also mapping between blocks and DataNodes will be generated [45].

Large HDFS instances run on a cluster of regular commodity computing devices that commonly spread across many racks. Typically, network bandwidth between machines in the same rack is greater than network bandwidth between machines in different racks.

Network traffic between different nodes within the same installation is more efficient than network traffic across installations. A NameNode tries to place replicas of a block on multiple installations to improve fault tolerance. However, HDFS allows administrators to decide on which installation a node belongs. So, each node knows its rack ID, making it rack aware.

One of good features of HDFS is its data awareness between NameNode and DataNode. JobTracker in NameNode schedules jobs to TaskTracker by knowing the location of data. This will reduce the traffic of the data across the network because NameNode avoids unnecessary data transfer. This has a great effect on performance, but when Hadoop is used in combination of other file systems, this feature is not available [23].

The block sizes in HDFS are big. It is typically 128 Megabyte to 256 Gigabyte for MapReduce applications and 1 Gigabyte or 2 for archival store. In HDFS there are mostly sequential reads and sequential writes and I/O operations on disks won't cause a bottleneck. There is also no coherency between readers and writers and it is very useful for scaling out. A client can read a file while it is being written [48].

*4.3 HDFS heartbeats*

There are several reasons that cause loss of connectivity between NameNodes and DataNodes. Therefore, each DataNode sends periodic heartbeat messages to its NameNode, and NameNode can detect loss of connectivity if it stops receiving them. The heartbeat interval is three seconds by default. The NameNode marks DataNodes not responding to heartbeats as dead and avoids sending further requests to them. Data stored on a dead node is no longer available to an HDFS client from that node, which is effectively removed from the system. If the death of a node causes the replication factor of data blocks to drop below their minimum specified value, the NameNode initiates additional replication to bring the replication factor back to a normalized state. Figure 10 depicts the HDFS process of sending heartbeat messages.

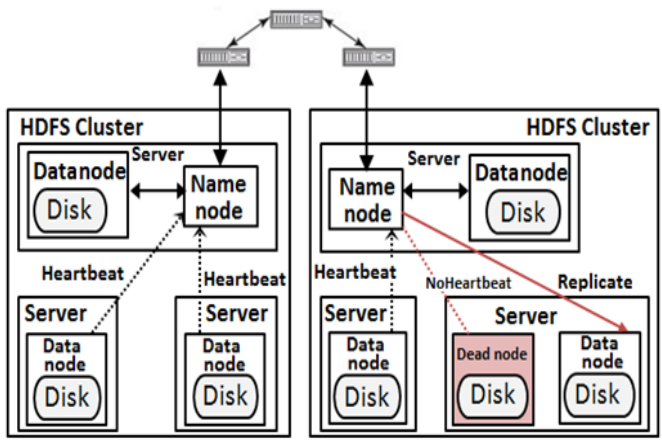

Fig 10: HDFS Heartbeat Process.

In order to synchronous Metadata updating in HDFS, there is a log file named Journal. The Journal is a write-ahead commit log for changes to the file system that must be persistent. For each client-initiated transaction, the change is recorded in the Journal, and the Journal file is flushed and synched before the change is committed to the HDFS client [40].

NameNode uses a transaction log called the EditLog to record every change that occurs to the file system Metadata like creating a new file or removing it. Changed replication factor of a file EditLog is stored in the NameNode's local file system. Entire file system namespace including mapping of blocks to files and file system properties is stored in a file named FsImage. FsImage is stored in NameNode's local file system.

FsImage and EditLog are central data structures of HDFS and Metadata about all data is recorded in them, so a corruption of these files can cause a HDFS instance to be non – functional. NameNode can be configured to keep multiple copies of the FsImage and EditLog. Multiple copies of the FsImage and EditLog files are updated synchronously [46].

There are three phases for Metadata replication and updating: initialization phase, replication phase and failover phase. In the first phase, multiple DataNodes register with the NameNode to take updated Metadata information, in the second phase, NameNode gathers Metadata from client request threads and sends them to DataNodes, so DataNodes answer with a heartbeat message to keep track of current status of Metadata, and in the last phase if NameNode doesn't send any heartbeat acknowledgement messages to DataNode, it means that NameNode failed and a leader election algorithm will be executed to select a new NameNode from remaining DataNodes [21].

*4.4 Checkpoint Node and Backup Node*

The NameNode in HDFS, not only has a role as serving client requests, but also can alternatively execute either of two other roles, one as a Checkpoint Node and second, as a Backup node. The role is specified at the node Startup; on a different host from the NameNode since it have the same memory requirements as the NameNode. It downloads the current checkpoint and Journal files from the NameNode, merges them locally, and returns the new checkpoint back to the NameNode [40].

The checkpoint file will never change by the NameNode. It will replace entirely by a new created checkpoint when requested by administrator or a checkpoint node during a restart. During Startup, the NameNode initializes the name space image from the checkpoint and then until the image is updated with the last file system state, it replies to changes from the Journal. Before NameNode starts to serve the client, a new checkpoint and an empty Journal to storing directories are written back.

One of HDFS features is backup nodes. Like check point node, backup node is able to create check points periodically. But in addition, backup node keeps an in memory, updated image from file name space which is always synchronized with the NameNode state. Backup node can be seen as a read – only NameNode. Backup node contains every Metadata of file system except the location of data blocks. It can do all operations of a regular NameNode except those which involve in correcting name space and or knowing the location of data blocks. Using the backup node provides the option of running the NameNode without persistent storage, delegating responsibility for the name space state persisting in a backup node.

To improve fault tolerance mechanism, NameNode can take advantages of secondary NameNode [49] and uses it for backing up data regularly and maintaining a protective mechanism[1]. The secondary NameNode is responsible for chekpointing the primary NameNode's persistent state [45].

HDFS executes a permission model based on the unix user/group model for files and directories which is similar to the portable operating system model (posix) [45]. For example each file and directory is associated to an owner and a group. HDFS permissions model supports read (r), write (w) and execute (x). Because there is no concept of file execution in HDFS, the x permission has different meanings. In another words, the x attribute indicates permission for accessing a child directory of a given parent directory. The owner of a file or a directory is the identity of a user who creates it. The group is the parent directory group.

*4.5 Hbase*

Traditional RDBMSs do not scale up well to thousands of users and their CPUs and I/O starts to make problem on the DB (data base). Hbase is an open source distributed column – oriented non – relational multi – dimensional data base that provides high – availability and high – performance. It creates a large table of properties on the HDFS which is named Hbase. Hbase uses an on – disk column storage format and has a key – based access mechanism to accessing a specific cell or a sequence of cells [50], [51], [52]. Hbase contains Java classes which allows it to be used as a resource or a sink for MapReduce jobs transparently [41], [53].

Hbase uses server side processing and reduces data transfer and distributes computing [54]. It is a distributed structured data table and uses a simple master – slave architecture which consists of a master server and a set of sub servers. Hbase is a sparse long – term storage on hardware and is multi – dimensional and ordered mapping table. The index of the table is row key – words, column key – words and time stamp. By using time stamp, multiple data can be stored in one column when data is uploaded [1]. Users can store series of data lines in Hbase table. Each line consists of sequenceable line key words, an optional time stamp and some (sparse) columns which possibly have data. All data are physically stored in Hbase and each table is divided into several subs – tables stored in appropriate places. Sub – table servers provide data services. Hbase uses only one master server to manage all subs – servers of the table and each sub – table server only contacts with just one master server which tells to each sub server to load sub – tables and provides services to them. Figurer 11 shows a simplified scheme of the Hbase architecture.

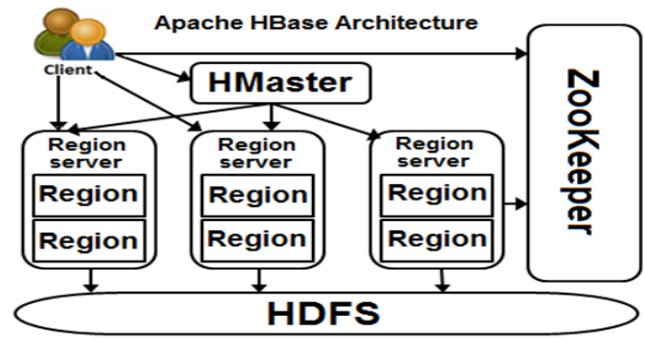

Fig 11: Apache Hbase Architecture.

An Hbase/Hadoop cluster includes three parts: HDFS for distributed file storage, Zookeeper which coordinates distributed services and Hbase that records lookups and updates in distributed file system. Nodes in Hbase cluster consist of several layers of software systems. Figure 12 – a depicts the layers of software in Hadoop. Each layer must perform correctly to be able to communicate across layers and nodes and entire system as shown in figure 12 – b [55].

Zookeeper is a distributed common consent engine which runs on a set of servers and maintains state consistency. It has concurrent access semantics and it provides producer/consumer queues, priority queues and multi – phase commit operations [39]. Zookeeper provides distributed configuration service, synchronization service and naming registry for large distributed systems.

Hadoop, Hbase and Zookeeper are based on Java SDK. The Hadoop layer is the basis of Hbase layer. Hbase does not need MapReduce, but applications of Hbase need it. Without Zookeeper, Hbase is not operational. Zookeeper tracks server failures and network partitions [55].

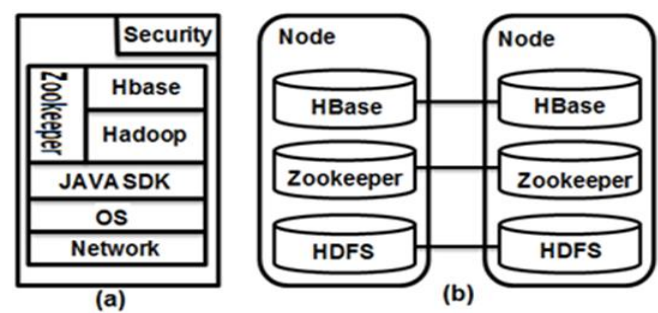

Fig 12: Layers of Software System in Hadoop.

*4.6 Hive, Pig and Flume*

Hive is a data warehouse framework built on top of Hadoop which simplifies "easy data summarization", "ad – hoc queries" and analysis Big Data sets that stored in Hadoop compatible file system. It is developed from Hbase by Facebook and uses a mechanism like SQL to save and retrieve data from Hadoop platform [1], [56]. Hive provides a mechanism to project structure into data and query the data using a SQL – like language called HiveQL. Users can manipulate data on the cluster by HiveQL [57]. Also Hive allows the usage of traditional MapReduce programs [58].

Hive is based on the MapReduce which is not suitable for low – latency queries. Hive "parses" HiveQL query, creates MapReduce program based on parsing results and then executes it. So Hive cannot be a replacement for traditional RDBMS systems. Its SQL – like language is used mostly for convenient and it's not a replacement for RDBMS. Figure 13 depicts a simplified look of Hive architecture.

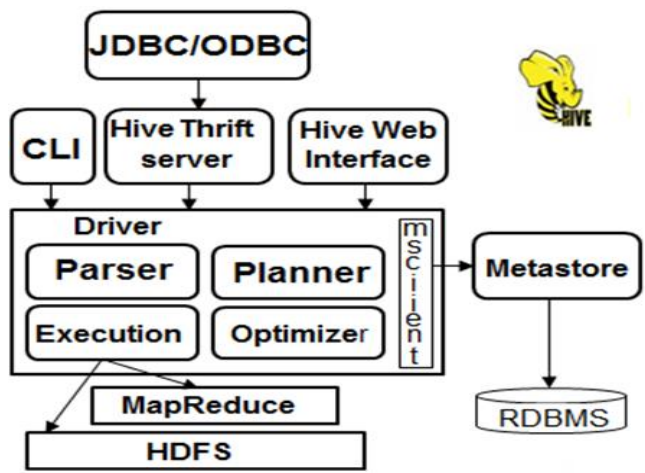

Fig 13: Apache Hive Architecture.

Pig is a framework with high – level data – flow language (Latin) which its compiler generates sequences of MapReduce programs to run in Hadoop. Pig is designed for analysing Big Data sets. It is originally designed for batch processing of data. Pig is a Java client side application. It has a compiler in its infrastructure layer which converts Pig Latin programs to MapReduce programs. Pig has an interactive shell named Grunt and does not change anything in Hadoop [39], [56].

Queries in Hive execute based on the MapReduce. Cloudera presented a new product named Impala to use with Hadoop. It is a SQL – like engine and provides fast interactive SQL queries. It bypasses MapReduce framework. By Impala business intelligent tools can run queries against data in HDFS and Hbase. It uses separate set of processes to bypass the MapReduce and access directly to HDFS and Hbase data [59].

Flume is a framework which provides reliable services to transferring and collecting Big Data. Flume architecture consists of three tiers: agent tier, collector tier and data store tier. A Flume agent installs on every data sources. It monitors logs such as console logs. Each agent has three components:

- Source: receives Flume events from external sources and delivers them to one or more channel.
- Channel: it is a temporary space which event data are stores in it until sink use them.
- Sink: it puts event data to other Flume sources or other external repositories.

An agent Flume collector gathers the events from other agents and stores them to an external repository like HDFS through its sink.

Flume provides end – to – end reliability because it guarantees the completion of storing data from source to destination. It also can guarantee that data will be send to next node in the presents of failures. It uses best effort policy to transferring data [34].

*4.7 YARN*

MapReduce has been overlooked completely in Hadoop 0.23 and now MapReduce version 2.0 (MRv2) or YARN is presented.

The main idea of MRv2 is that it separates the two major functionalities of JobTracker which is resource management and scheduler/monitoring into separate daemons. The idea is that there can be a global resource manager (RM) and an Application Master (AM) for each application. An application is either a single job like classical MapReduce jobs or a DAG (Directed Acyclic Graph) of jobs.

The data – computation framework is contained form resource manager, per – node slave and the node manager (NM). The resource manager is the ultimate authority that distributes resources among all applications in the system.

The per – application master is a special library that negotiates for resources with resource manager. It also works with node managers to executing and monitoring tasks.

The resource manager has two major components: scheduler and application manager.

The application manager is in charge of accepting jobs and submissions and negotiates with the first container for running the application and it provides a service for restarting the application master container on failure.

Node manager is a framework agent per machines which is responsible for containers and monitors their resource usage and reports to the resource manager / scheduler.

Per – application manager is responsible for negotiating for appropriate resource containers from scheduler and tracking and monitoring their status.

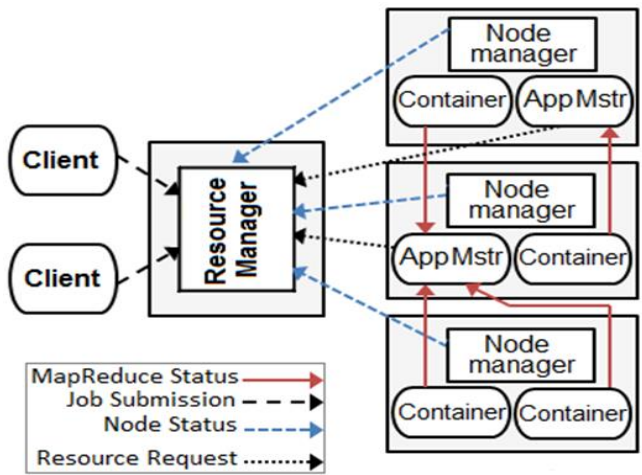

Fig 14: Scheduler Overview in Hadoop.

MRv2 maintains API compatibility with previous stable releases. This means that all MapReduce jobs should still run on top of MRv2 without any changes with just a recompile. In the Figure 14 we have shown an overview of job scheduling (which we discuss later) in Hadoop.

The lifecycle of a job in YARN is as follows:[31]

- The Job Client submits a job to the YARN application manager (AM)
- The YARN RM – AsM (Resource Manager – Applications Manager) communicates with container for MapReduce application master and lunches the MR AM (MapReduce Application Manager) for the job.
- The MR AM registers with RM – AsM.
- The Job Client maintains information about the MR AM and then it communicates to AM for status, counters and etc.
- The MapReduce application manager determines how input should be divided and constructs resources requests and sends them to scheduler.
- The Job setup APIs of the Hadoop for MapReduce OutputCommitter start to running.
- The MapReduce application manager submits the resource requests to the scheduler and gets a container from RM, and then it schedules tasks on the obtained container.
- The MapReduce application manager monitors all tasks to complete successfully, if a task fails or stop responding, then it requests to alternate resources for the failed job.
- After a task completes, MapReduce application manager runs appropriate task Cleanup code of Hadoop MapReduce OutputCommitter.
- When all tasks are finished, MapReduce application manager runs the job commit APIs of Hadoop MapReduce OutputCommitter.
- Then MapReduce application manager exists.

5. HDFS Failures and Optimization

Although Hadoop is a fault tolerance system, several issues and errors can occur in it. In this Section we illustrate four top failures that might occur in HDFS and also describe some optimizations in it.

5.1 Hardware and Software failures

HDFS may consist of thousands of server machines, each of them stores a part of file system. Since there are a lot of components and each component has a probability of failure or malfunctioning, therefore some components might be non – functional. Detecting failures and a quick automatic recovery from them is the core architecture goal of HDFS. MapReduce is very memory – bound and so it is inclined to misconfigurations.

The common failures in HDFS are: NameNode failure, DataNode failure and network partitioning. (Network faults result in nodes to fail) [60].

Each DataNode sends heartbeat massages to NameNode periodically. Network partitioning may cause a sub set of DataNodes to loss their connection to NameNode. NameNode detects this situation by absence of the heartbeat massages and marks the DataNode without heartbeat massages as a dead node and does not send any I/O request for it. Any data registered in these dead nodes is not available in HDFS. Also the death of DataNodes may cause the replication factor of some blocks drops bellow of their specified value. The NameNode constantly tracks which blocks need to be replicated and initiates replication whenever it's necessary[46].

Sometimes there might be a bug in the software however having replicated data reduces the impact of this problems [45].

5.2 Cluster rebalancing and Data block rebalancing

In Hadoop system, nodes simultaneously play both computing and storage rules. A file splits into several pieces and distributes across distinct nodes. In this environment any node or file can be deleted, upgraded, replaced or added to the system dynamically. These can results in load imbalance of file system and since mapping data and tasks is on NameNode, NameNode will become a single point of failure. In other words, in Hadoop system, NameNode can be a Bottleneck of load balancing [6]. The HDFS architecture is compatible with data rebalancing scheme. A scheme might automatically move data from one node to other nodes when the free space on a DataNode drops below a threshold. In the event of suddenly demanding a particular file, this scheme creates additional replicas dynamically and rebalances other DataNodes in the cluster. These types of data rebalancing are not implemented yet. But there are some approaches for load rebalancing in distributed file systems that can be applied to Hadoop.

*5.3 Data integrity*

The data fetched from a DataNode might arrive corrupted. This corruption may occur because of failure of storage device, network failure or a software bug or any other possible reasons. HDFS client software implements a checksum for checking on the content of HDFS files. When a client creates a HDFS file, it computes a checksum for each block of file and stores it in a separate hidden file in the same HDFS name space. When a client retrieves the file contents, it verifies if the data that receives from each DataNode and stored with checksum matches the checksum stored in the file associated checksum file and if not, then the client can retrieve the block from another DataNodes which has a replica of that block [46], [47].

*5.4 Metadata Disk Failure*

Any update to either the FsImage or EditLog causes each of the FsImages and EditLogs to get updated synchronously. This synchronous updating of multiple copies of the FsImage and EditLog may degrade the rate of namespace transactions per second that a NameNode can support. However, this degradation is acceptable because even though HDFS applications are very data intensive in nature, they are not Metadata intensive. When a NameNode restarts, it selects the latest consistent FsImage and EditLog to use.[44]

*5.5 Optimizing HDFS*

The performance of MapReduce and Hadoop particularly is underlying file system that can have an important impact on the overall MapReduce framework. Optimizing HDFS will increase the efficiency of the MapReduce application in Hadoop. The poor performance of HDFS can be because of existing challenges of maintaining its portability, including disk scheduling under concurrent workloads, file system allocation, and file system page cache overhead. HDFS performance can be improved through the use of application – level I/O scheduling while preserving portability. More improvements are possible through decreasing fragmentation and cache overhead, at the expense of the portability.

The performance of Hadoop and HDFS depends on several metrics. Rasooli et al introduced five top metrics that can be measured as follows:

- Average completion time: the average time to complete all jobs
- Dissatis faction : a metric to determining the satisfaction of the minimum share requirements of the user
- Fairness: dividing resources among applications and users fairly
- Locality: clarify that how many tasks are running on resources have their stored data. In processing BigData, this is a big deal
- Scheduling time: it is the time that scheduler spends to schedule all incoming jobs.

These metrics have trade – off with each other and improving and optimizing one metric can cause other metrics to drop [61].

Some parts of Hadoop can be optimized particularly. For example Hive can be optimized by converting common – joins into Map – joins. This can be useful when carrying out join operations on the cluster. Currently Hive in order of determining if a common – join should be converted into Map – join, uses the size of tables on disk. Experiments show that this is a traditional decision criterion which does not succeed to identify optimization opportunities close to the decision frontier [57].

Also for using MapReduce most effectively for solving a problem, there are some algorithm characteristics which are necessary for effectively and beneficially implementation in MapReduce. These characteristics are presented as rules bellow:

- Outputs are a list of values or the problem requires a list of intermediate values
- Each value can be computed by Reduce function using just a subset of input data
- The numbers of Reduce steps are greater than or equal to the number of available processing nodes
- The data and information that are needed for computing output shouldn't have a joint hierarchical relationship among the input and the results.
- Data required for each Map and Reduce function should fits within available processor node storages.

Each reduce step, reduces the number of data objects by an order of magnitude or even more [28]. As mentioned before, between a set of Map and Reduce tasks, some intermediate data is transferred from the Map tasks to the Reduce tasks through the network interconnect of a cluster. This communication between Map and Reduce tasks is named shuffle phase of a Hadoop application. The execution time of an application in Hadoop is greatly affected by the amount of data transferred during shuffle phase. Improving performance of the shuffling phase for shuffle intensive applications (applications which have a lot of intermediate data and shuffling phase) is very critical. In [29] it has been illustrated that pre – shuffling enabled Hadoop clusters are faster than default Hadoop clusters.

We will discuss Hadoop scheduler later, but we mention an approach which optimize scheduler and increase the performance of Hadoop that is showed in [62]. They purpose a predictive scheduling algorithm to arrange tasks of MapReduce and manage the loading by preloading the data from local disks to caches as late as possible without delaying on lunching arriving tasks. Hadoop by default assign tasks randomly to nodes and loads data from disks whenever it is needed. CPU's of the processing nodes will not process new tasks until all the input data resources are loaded into the nodes main memory. The coordination between CPUs and disks for data I/O has a negative impact on performance. This new algorithm helps Hadoop cluster to preload data that will be needed before lunching tasks on DataNode and by this, the results showed that they improved the performance of MapReduce, shortened the waiting period and preloading the data earlier than the new task assigned.

There are also some approaches to improve the performance of Hadoop (not just HDFS). One of these approaches is Hadoop++. Hadoop's processing pipeline has about ten User Defined Functions (UDFs) like Map function and Reduce function. These functions can be used to placing any code inside the processing pipeline. In [63] they placed indexing and co – partitioning algorithm into Hadoop. For example they changed the group() UDF to control grouping and shuffle() UDF to control shuffling phase. By this they could create separate indexes for every HDFS block. Their experimental results showed that Hadoop++ is more than 20 times faster than Hadoop.

The shuffle and sort phases sometimes overfill the top of switches in the network especially when Hadoop is implemented in a multi – rack cluster. Also MapReduce tasks often have hot spots which mean a computation is taking more time because of insufficient bandwidth of some nodes. There is an approach named Hadoop – OFE (an OpenFlow enabled version of Hadoop) which enables an application to adjust the network topology as needed by the computation. Hadoop – OFE dynamically manipulates the network topology to improve performance of Hadoop[64].

There are also some other optimizations in Hadoop MapReduce and HDFS. For example, Jayalath et al introduced a model named G-MR which is a system for executing MapReduce jobs on geo – distributed data sets. Their model improves processing time and cost for geo – distributed data sets [2]. Zaharia et al, improves the performance of Hadoop by making it aware of the heterogeneity of the network [65]. MapReduce Online is a development of MapReduce which allows HDFS components to start execution before the whole data become materialized, for example, if there are just some parts of Map function that are completed, then Reduce function can begin processing them [66]. Chang et al designed an approximation algorithm that sorts the MapReduce tasks and minimizes the overall execution time of jobs [67]. Yang et al attached an extra phase to MapReduce phase named merge that follows Map phase and merges multiple reducer outputs from different lineages. Their model works well for heterogeneous data [68].

6. Hadoop Scheduler

In order to improve performance in Hadoop, multiple jobs operating on a common data file can be processed as a batch to eliminate the overhead of redundant scanning. In practice, jobs often do not arrive at the same time, in batching it means longer waiting time for jobs that arrived sooner.[69]

To support enter – job parallelism, should be focused on job scheduling mechanism based on shared resources. In particular, there are two kinds of sharing resources in MapReduce framework. The first: multiple jobs can use shared computing resources like CPU time, memory and disk in a MapReduce cluster.

The second: inter – job parallelism is exploited by taking advantage of the common processing that may be shared by several jobs. For example if multiple jobs access to the same file, then it is suitable to access file once for all such jobs.

A scheduler, associate MapReduce tasks to Hadoop resources. Kamal and Kemafor extended real time cluster scheduler to derive minimum map and reduce task count to performing task scheduling which meet the deadlines. They designed and implemented a constraint scheduler for Hadoop to fulfil this goal. Their experimental results showed that when deadline for jobs are different, constraint scheduler assigns different number of jobs to TaskTracker to make sure that all of them will be completed in the constraint time [70] .

When a job is submitted, the scheduler should determine if this job will be completed in the specified deadline or not. The scheduler concentrates on the number of free and available slots at the given time or later in the future. Therefore, in order to make sure that all jobs will be finished in the given time, the right number of tasks should be assigned to TaskTracker. Some strategies for assigning certain amount of tasks are:

- Assigning the whole MapReduce tasks: if the number of tasks is less than available slots, then assign all tasks, else, assign tasks to all available slots in the cluster. This may cause jobs that submitted later, won't have enough slots to run.
- Assigning minimum tasks: only assign a minimum number of tasks that should be finished for a job to be completed in a given time. In this strategy, free slots are available for jobs that will submit later.
- Assigning a constant number of tasks.

A job is schedulable when the number of free slots is equal or bigger than the minimum number of tasks for map and reduces.

In Hadoop, pluggable schedulers are supported. Hadoop config file should be modified to use. The goals of designing the scheduler are:

- It should be able to give users a feedback whether a job can be completed in a given time or cannot be completed in the deadline and proceed to complete the job, and if a job cannot be completed in the deadline, users should have the option to resubmit the job with modified deadline requirements.
- It should maximize the number of jobs that can run on a cluster in a given time.

Previously, scheduling tasks in Hadoop were limited to FIFO, Fair scheduler and Capacity scheduler methods, recently, research on Dynamic Proportional Scheduler [71] provides more job sharing and prioritization capability in scheduling. This resulted in increasing sharing of cluster resources and also increasing the differentiation in service levels of various jobs. By focusing on minimizing the total completion time of a set of MapReduce jobs, time estimation and optimization for Hadoop jobs has been considered. Most efforts in scheduling are on handling various priorities of jobs and most efforts in time estimation efforts are based on estimating the run time of already running jobs. Deadlines for jobs are different, so the scheduler assigns different number of tasks to TaskTracker and makes sure that the specified deadline for all jobs will be met.

There is a genetic algorithm for task assignment in environments that are based on Grid. The scheduling difficulties in these environments are the mapping of tasks onto a suitable service level in order to minimize the execution time of a workflow and complete it in the given deadline.

Existing Map Reduce Schedulers that are based on resource utilization can be classified into two categories:[72]

- Full utilization which maximize the use of resources.
- Partial utilization which enable concurrent processing.

Hadoop default scheduler uses first class of above. FIFO scheduler is available in all Hadoop implementations [73]. This scheduler manages a FIFO queue for all submitted jobs according to their submission time or priority that specified by users. All jobs are sorted first based on their priority and second on their submission time. This method allows a job to take all slots to do its tasks and other jobs cannot use the cluster until the current job is finished. This policy is very restricted and increase the total completion time. For example, when a job is waiting for its last map task to complete, despite that other maps are done, the reduce task cannot start until all map tasks are finished. This happens because slots assigned to map tasks are not released until all these map tasks are done. Jobs that are ahead in the queue, block jobs that arrive in a later time or have a lower priority. If the number of given jobs is large, FIFO scheduler will cause a great delay and can't guarantee that jobs will be completed in a deadline.

## 6.1 Scheduling methods

In general, the default scheduler in Hadoop is FIFO scheduler which is available in all Hadoop versions. But there are new approaches that are trying to improve scheduling performance and consequently optimize Hadoop. Some of them are implemented in new versions of Hadoop. In this Section we introduce several approaches for implementing scheduler in Hadoop.

### 6.1.1 A new policy in FIFO

If a map task is finished, the slot of this map task is released and the output of this map task is sent to reduce task. This new policy decreases the total completion time.

Some other schedulers have been developed to adopt partial utilization of resources. For example yahoo uses capacity scheduler and Facebook uses fair scheduler.

Capacity scheduler manages multiple queues which a fraction of resources in the cluster are allocated to each of them. These fractions are independent and so more jobs can be executed concurrently.

Fair scheduler organizes jobs into multiple user pools. Each pool shares whole resources fairly. All running jobs can receive a fair share of resources and more jobs can be executed in a given time. Two disadvantages of this scheduling method are:

- Since each job gets less resources, its execution time will be longer
- Still each job is running independently.

As mentioned before, Hadoop by default uses FIFO scheduler, which is a non – preemptive scheduler, however preemptive scheduling had some advantages in comparison to such as total completion time and slot utilizations [71]. There are some approaches which consider scheduling problems with preemptive solutions under deadline constraints. Their results indicated that preemptive scheduling is much more efficient than non – preemptive scheduling under deadline constraints [72].

### 6.1.2 The job queue manager (JQM)

The job queue manager (JQM) is a component that handles sub – jobs. It keeps a job queue that holds sub – jobs of existing jobs. When a new job arrives, instead of being submitted to MapReduce directly and immediately, it first divides into sub – jobs and then stores in the job queue. JQM analyse the sub – jobs and then align them with other existing sub – jobs in the job queue, so all sub – jobs that access the same segment will be batched together. JQM organizes these batches according to the information retrieved from MapReduce framework and dynamically determines the set of sub – jobs to batch processing. When all batches of sub – jobs that exist in the cluster have completed, JQM launches the batch of sub – jobs that are ahead in the queue to MapReduce [72].

### 6.1.3 Scheduling heterogeneous nodes

A scheduler assigns the MapReduce tasks to Hadoop resources. The considerable challenge is the growing number of both tasks and resources in a scalable manner. The heterogeneous nature of nodes in the Hadoop system increases this challenge [61].

The COSHH (Classification and Optimization based Scheduler for Heterogeneous Hadoop) scheduler is a Hadoop scheduler which is designed to consider system and user heterogeneity and make decisions by using this information. So COSHH considers incoming jobs and classify them and then finds a matching class of the resources based on this information which contains the requirement of the job classes and features of the resources. COSHH algorithm solves a linear programming problem (LP) to find a sufficient matching of jobs and resources [71].

The COSHH algorithm uses the set of suggested job classes for each resource, and considers the priority of jobs, required minimum share of resources, and fair share of users to make a scheduling decision.

The comparison of three methods:

Jobs have different executing time. COSHH algorithm matches job classes to the resources considering heterogeneity; FIFO does not take to account job sizes, this cause a problem that small jobs might have to wait behind the large jobs for a long time. Fair scheduling and COSHH algorithm does not suffer from this problem. In the fair scheduling, jobs wait in different pools based on their size and resources are allocated to each pool in a fairly manner. So in the fair sharing algorithm, different sized jobs are executed in parallel. The COSHH algorithm assigns the jobs to the resources based on the job size and the execution rate of the resource, so this problem does not exist in this algorithm too.

*6.1.4 S3: a novel shared scan scheduler for Hadoop*
S3 shares the scanning of a common file for multiple jobs. This scheduling is designed for jobs that arrive at different times and its goal is process jobs as soon as possible. S3 divides a job to several subs – jobs and a sub – job contains the exact amount of work that utilizes the whole resources of the cluster for one round of execution. It means that the number of sub – jobs corresponds to the number of rounds that is required for the original job to complete [69].

*6.2 YARN schedulers*

The next generation of schedulers in Hadoop is YARN schedulers. The scheduler in Yarn is responsible for allocating resources to various running applications and it does this by considering constraints of capacity and queue and etc. This is a pure scheduler because it does not any monitoring or tracking of status of application and it will not guarantee restarting failed tasks due to hardware failure or even application failure. The scheduler performs its tasks based on the application requirements of resources. The scheduler executes based on the abstract notion of resource container that includes resources such as memory and CPU and disk and etc. In previous versions of Hadoop only memory was supported.
The scheduler has a pluggable policy (plug in) which is responsible to divide cluster resources among applications or queues. The capacity scheduler and fair scheduler are two examples of this plug in.

The capacity scheduler supports hierarchical queues to allow more predictable resource sharing in the cluster.

*6.2.1 Fair scheduler*
Fair scheduler is a pluggable scheduler for Hadoop that provides a way to sharing large clusters. The implementation of fair scheduling is under development and must be considered experimental [74].

Fair scheduling is a way that resources allocated to every application such that every application gets an equal average share of resources. The next generation of Hadoop is capable of scheduling multiple resources like CPU and memory. Currently Hadoop just supports the memory. So a "cluster share "is a proportion of aggregated memory in the cluster. When just one application is running, it uses entire cluster. When new applications assigned to the cluster, free resources are assigned to them, such that each application gets almost the same amount of resources. Unlike the default Hadoop scheduler which performs a queue of applications, this scheduler allows small applications to complete in a short time and don't starve from long lived applications. Also this is a reasonable way to share a cluster among multiple users. Fair scheduling can work with application priorities. The priorities can be used as weights to specify the fraction of total resources that an application can get.

Fair scheduler organizes applications into queues and shares resources among these queues fairly. By default, all users share the same queue which is called "default". If an application list and a queue in a resource container request, the request will be submitted to that queue. Also it is possible to assign queues based on user name which is included with the request through the configuration. Fair sharing is used within the each queue to share the capacities among the running applications. It is also possible that assign weights to queues in order to share the cluster non – proportionally in the config file.

Fair scheduler supports hierarchical queues. All queues descend from a queue which is named "root". The usable resources are shared among the children of the root queue fairly. Then these children distribute the resources assigned to them among their own children. Applications might only be scheduled on leaf queues. Queues can be specified as children of other queues and this is possible by placing them as sub – elements of their parents in the scheduler config file.

By default, fair scheduling allows all applications execute, but it is possible to limit the number of running applications per user through the scheduling configuration file. It is useful when a user suddenly submits hundreds of applications or to improve the performance when too much applications at once would cause too much intermediate data to be created or too much context – switch. Limiting the number of running applications does not cause any application that is submitted subsequently to fail; they just remain in the scheduler queue until some of applications that assigned earlier, finish. The applications from user or queue are sorted and chosen in order of priority and submit time, just like the default FIFO scheduler in the Hadoop.

In addition of fair sharing, fair scheduler allows assigning guaranteed minimum shares among the queues to ensure that certain users, group or applications can always get appropriate amount of resources. When a queue ID contained application, it can get at least the minimum share of resources. But when a queue does not need its entire share, the extra will be shared among other queues. When this queue does not contain applications, this allows the scheduler to guarantee the queue capacity while utilizing resources efficiently.

*6.2.2 Capacity scheduler*

Capacity scheduler is a pluggable scheduler for Hadoop that allows to multiple clients to share a large cluster, such that in a given time, under resource allocation limits, resources assign to their applications [71], [75].

The goal of designing this scheduler is that Hadoop applications which are shared among several users in a cluster execute in a friendly manner and maximize throughput and utilization of the cluster. Each organization has its own private computing resources; capacity scheduler is designed to allow large clusters to be shared while guarantees capacity for each organization. The main idea is to share usable resources among the multiple organizations which can find their computing requests in the cluster. So each organization can access resources that others do not use them, this advantage will cause the flexibility in their costs.

Sharing resources among multiple organizations needs a lot of supports, because it is necessary to provide capacity and security for each organization in order to be sure that is impermeable for a big corrupt program or a user or it configures what is belong to itself. Capacity scheduler provides rigorous restrictions to be sure that a program or a user or a queue cannot use inappropriate amount of cluster resources. Also, it provides limitations on initialized/pending applications from a user or a queue to ensure fairness and stability of cluster.

To provide further control and predictability on resource sharing, this scheduler supports hierarchical queues to ensure that resources will be shared among the sub – queues of an organization before other queues are allowed to use free resources and capacity scheduler do this by providing affinity for sharing free resources among applications of a given organization.

It supports several features including:

- Hierarchical queues: as we described it in above.
- Capacity guarantees: queues are assigned a fraction of capacity of the Grid, so they have a certain amount of resources of cluster. All applications submitted in a queue can access the capacity allocated to the queue. Also administrators can configure soft limits or even hard limits on allocated capacity to each queue.
- Security: each queue has a specific ACL (Access Control Limit) which controls which users can submit their applications in the queue and also there are security guards to ensure that users cannot see or modify applications from other queues and also it supports per – queue system administrator roles.
- Elasticity: free resources can be allocated to any queue beyond its capacity. When there is a request for these resources from queues running below capacity in a future point and tasks assigned to these resources are complete, they will be allocated to those queues that need them. But in this scheduler, preemption is not supported. This will result that resources are predictably and classically accessible for queues, so the utilization will be improved.
- Multi – tenancy: a comprehensive set of limitations are provided to prevent from one application or queue get all the cluster or queue resources. This ensures that cluster won't be drowned.
- Runtime configuration: a console is provided for users and administrators and they can observe resources currently assigned to system queues and modify queue definitions like capacity and ACLs in a secure manner to prevent application disruption as much as possible. Also it is possible to add additional queues at runtime, but it's not possible to delete them.
- Drain applications: administrators can stop queues at runtime to ensure that while current applications are running to be completed, other applications cannot submit to the queue. If a queue is in stop state, new applications cannot submit in itself or in its children. Existing applications continue to completion, so the queue can be drained, also administrators can start a stopped queue at any time.
- Resource based scheduling: this scheduling supports resource intensive applications. When an application need more resources than default, scheduler can support it by accommodate application with differing resource requirements.

Fair scheduler and capacity scheduler don't consider deadlines that users define for jobs. To implementing deadline scheduling in Hadoop, data locality, predictability of the execution time of tasks and the preemption are the important factors that must be considered [76].

7. Hadoop on Cloud, Grid and GPU

The main goal of Hadoop existence is dealing with Big Data efficiently. A Hadoop cluster is made of commodity computers which provide a processing power. Since it is close to other distributed systems like Cloud or Grid, it makes sense to implement Hadoop in such systems and take advantage of using its combination by them.

*7.1 Hadoop – based Cloud Computing*

Hadoop is an open source software platform for distributed computing which is capable of parallel processing on a massive amount of data and widely uses the Cloud computing for its purposes.

In Cloud computing, main system body consists of virtual servers that provide services simultaneously for clients all around the world. A useful Cloud computing model is presented with a set of Cloud infrastructure services to alternate the traditional servers and computational clusters.

Hadoop on Cloud is a Hadoop cluster created by demand and is consisted of several virtual machines from multiple Cloud sites which are connected together through a broadband network links across the Clouds [41], [77].

*7.1.1 Cloud computing architecture of Hadoop*

Figure 15 shows the Cloud computing architecture of Hadoop. Hadoop is not only a distributed file system for storage, but also is a framework which can execute distributed applications on a large cluster composed of regular computer hardware. In order to provide the reliability, HDFS produce multiple replicated data blocks and places them on computers of server cluster. MapReduce deals with data stored on nodes.

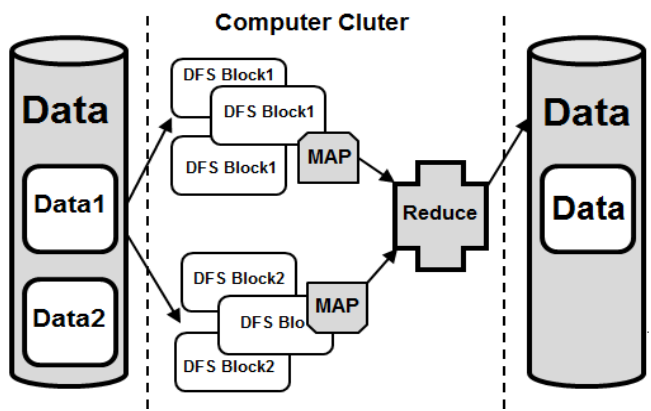

Fig 15: Hadoop Architecture overview in Cloud

*7.1.2 Cloud computing model of Hadoop*

As shown in Figure 16, Hadoop cluster uses a master – slave structure. Master is in charge of NameNode and JobTracker. The main due of JobTracker is initiating tasks, tracking and distributing implementations and executing them. Slave is responsible for DataNode and TaskTracker. TaskTracker manages local data processing and reducing data results according to requests of applications and reports status and performance to JobTracker. NameNode and DataNode are responsible of all HDFS tasks while JobTracker and TaskTracker deal with MapReduce tasks [78].

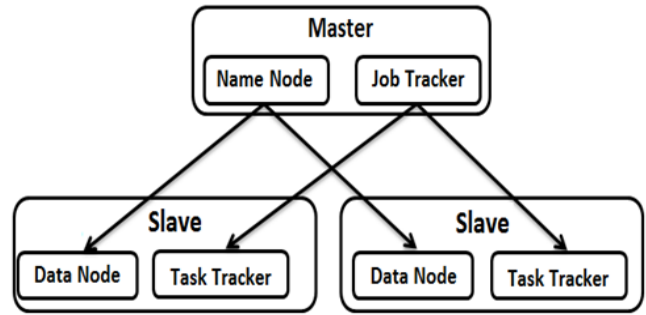

Fig 16: Master/Slave Structure of Hadoop Cluster.

Hadoop opened a new door to Cloud computing by MapReduce. The Hadoop system for data intensive cluster computing based on MapReduce has been adopted by Yahoo, Facebook, Google and Amazon and is broadening rapidly. Hadoop clusters are configured on demand from shared resources and infrastructures. Hadoop and Cloud platform combination has been developed for large amount distributed data processing for users and industry. Data sets and Cloud clusters are used by different organizations and users, so some works has been done to enable building Hadoop clusters on networked Clouds.

Instead of placing a repository on the Cloud site for a computation, it would be better to be able to distribute the computation to match the placement of data and resources. This goal can be achieved by networked Cloud infrastructure services, since these services can make it possible to control the location of virtually provisioned infrastructure and also by these services users can configure network topologies to link virtual resources on demand. Hadoop needs high bandwidth when operating on large data sets; however it is insensitive to transit latencies. The combination of latency tolerance in Hadoop and optional bandwidth allocation in networked Clouds make it practical to run Hadoop workflow across distributed Clouds which are connected through high bandwidth circuit networks.

MapReduce distributed data processing architecture is an excellent tool for data – intensive analysis on Clouds and in regular general clusters, since Hadoop has a very strong fault tolerance features and has the scalability and it's easy to use [41], [79].

There are multiple options to use MapReduce in Cloud environments, like using MapReduce as a service, Configuring a MapReduce cluster on Cloud or using Cloud MapReduce runtimes. One of these runtimes is Azure MapReduce which uses the Cloud infrastructure services. Cloud environments are not as stable as traditional computing clusters, so Cloud applications should be developed to predict and prevent these failures [80].

Azure MapReduce is a decentralized and distributed run time for windows operating systems which is developed by using Azure Cloud infrastructure services. Using these services allows implementation of Azure MapReduce to take advantage of provided Clouds services like scalability and high availability and since these services are distributed naturally, the single point of failure and bottleneck in networks and storage bottlenecks and overhead of managing these services are handled [80].

Azure map –reduce has overcome the access problems by reviewing the system design, it is not depend on immediate access to data across the all worker nodes[81].

The scientific MapReduce applications executed on Cloud infrastructure have better performance and efficiency in compare to MapReduce applications that executed on traditional clusters. The features such as horizontal scalability, ease of use for programming model makes using MapReduce a very excellent option for computational scientists and since it can use commodity computers, it can be handy when in – house computer clusters are not available. The oscillation in MapReduce performance on Cloud in a long time period is minimal and it shows the consistency and predictability of application performance in a Cloud environment.

The usage of Cloud services usually takes a longer latency than optimized non – Cloud computers and it does not guarantee the first time access to data. To overcome these issues, coarser grained MapReduce tasks can be used.

Cloud infrastructure services moves from static arrangement of resources which last for a long period of time toward a dynamic arrangement which resources can be shared and allocated on demand. These Clouds provide virtualized infrastructure as a unified hosting substrate for various multiple applications. There are several technologies which providing resources according to the need, such as Amazon Elastic Compute Cloud (EC2), Eucalyptus, Nimbus, and IBM's Blue Cloud [78]. In the Section 11 we discuss the challenge of optimizing Hadoop on Cloud infrastructure.

These failures are common in using Hadoop in the large industrial Clouds:

- Machine failure: it is a usual system error in distributed computing and can be solved by manually turn off the failed machine in the cluster.
- Losing the supporting library: while extending the cluster, new machines in the cluster may miss supporting libraries or the version of supporting libraries on them may be outdated. To solve this problem, a required library of the analysis can be removed.
- Lack of disk space: when a BigData application is executing on the platform, it may cause the disk space shortage due to large amount of intermediate data [30].

Virtual topology (VT) requests are the requests for embedding resource topologies. The VT requests are expressed by and extended variant of network description languages. They determine virtual edge resources and a topology to interconnect them. Depending on if nodes belong to a specific Cloud site or whether the system is free to choose them, the requests can be bound or unbound.

### 7.1.3 Distributed MapReduce on the Grid

Gris computing is a distributed network computing system that coordinates network resources and gains a common computational objective. Grid computing is usually used for computation – intensive applications (mostly scientific applications) [23].

Hadoop HDFS has features like low cast of deployment, maintenance and scalability. The scalability and stability of HDFS into the Map of Grid computing is very beneficial. Hadoop is adaptive to new hardware and varying requirements and is capable of being integrated and interfaced with various gird middleware easily [82].

There is a Hadoop framework executable on the open science Grid (OSG) and Hadoop on the Grid (HOG) is based on the OSG. This method is different from previous MapReduce platforms that run on the environments like Grid or Clouds. HOG provides a free dynamic and flexible MapReduce environment on opportunistic resources on Grid. Also it can provide free scalable services for users who want use Hadoop MapReduce [42].

HOG improved Hadoop fault tolerance by mapping data centers to virtual racks and creating multi – institution failure domains. All modifications in HOG are transparent to existing Hadoop applications. HOG has a rapid scalability and provide a comparable performance to a specific Hadoop cluster.

In OSG users that do not have resources execute their applications opportunistically, because they can pre-empt when a resource can allocate to them. Processing job over dedicated time or a need for resource from a machine can cause preemption on remote OSG site. OSG users cannot control preemption policies determined by the remote site.

Hadoop's fault tolerance considers two failure levels and data replications to avoid losing data. The first is node level. In this level when a node fails, data integrity of the cluster should remain and the second is rack level, when all nodes in a rack fails, data still should remain safe. In HOG there is another level: site level which means when a whole site fails, HOG's data placement and replication policy handle the failure and places data blocks.

OSG is a national organization which provides services and software for a distributed network of clusters. OSG user has a personal certificate which is verified by a virtual organization. A virtual organization is a set of organizations which is a platform for computational researches and sharing resources. A user receives a X.509 certificate and uses it for exploring remote resources in the virtual organization.

As shown in Figure 17, HOG architecture is consists of three parts: the first part is Grid sub – mission and its execution. In this part the requests form Hadoop worker nodes are sent out to the Grid and their execution will be managed. The second part is HDFS which is running across the entire Grid and the third part is MapReduce framework which executes MapReduce applications across the whole Grid. When HOG starts to work, a user can request the Hadoop worker nodes to run on the Grid. The number of worker nodes can be elastically increase or decrease by submitting or removing the worker node jobs. The Hadoop worker node is consists of both DataNode and TaskTracker node. The sub – mission file defines the job properties of Hadoop worker nodes.

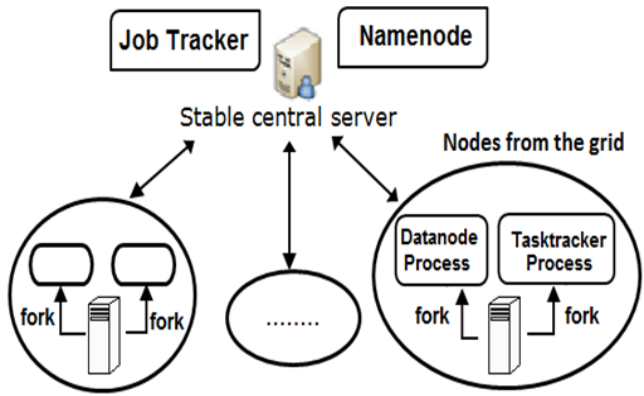

Fig 17: HOG Architecture.

The Hadoop instance running on the Grid has two major components: HDFS and MapReduce. Master servers which are NameNodes for HDFS and JobTracker for MapReduce are the single point of failure for HOG system. So they reside on a stable central server. If master server becomes unavailable, MapReduce jobs would be stopped and HDFS would be unavailable. But data would not be lost. When Grid jobs starts, slave servers will report to the only master server.

In Hadoop, worker nodes must be available and reachable for each other. Some clusters in the OSG can be behind one or more machines that uses network translation areas (NAT) service. Hadoop cannot communicate with those nodes behind a remote NAT, because NAT blocks direct access. So we are restricted to the sites on the OSG which have public IPs on their worker nodes.

Failure in the Grid is common and it is the result of site unavailability or resource preemptions. This is important that HOG rapidly covers lost data by data redistribution and processing on the other remaining nodes and request more nodes from Grid. Hadoop master node receives heartbeat messages from worker nodes periodically which are a report of their health. In HOG the time between heartbeat massages is decreased. Also the timeout time for a worker node is decreased too. If worker node does not report after 30 seconds, this node would be marked as a dead node for both NameNode and JobTracker.

Creating and maintaining a distributed file system on a heterogeneous set of hardware and resources is a challenging job to do. Hadoop is designed to manage this challenge suitably for frequent failures. In Hadoop DataNodes will contact the NameNode and report their status like information on the disk size on remote DataNode and available amount of space for storage for Hadoop. NameNode determines which data files should store on DataNode by location of the node using rack awareness and by percent of space which Hadoop can use from Grid.

The goal of implementing HOG is to provide a Hadoop platform for users to run on the Grid. They shouldn't be forced to change their MapReduce codes to adopt with Hadoop. So in HOG there are no changes in the API to MapReduce. When TaskTracker contacts JobTracker in the central server, JobTracker marks the available nodes for processing. TaskTrackers report their status to the JobTracker and accept task assignments from it. A task has at most two copies of execution in the system at any time.

In HOG, increasing the number of nodes will not decrease the response time. It has several reasons: HOG is based on the dynamic environment and worker nodes are opportunistic, when some nodes leave, HOG automatically request more nodes from OSG to recompose the lack of missing worker nodes. This will take some time for requesting, configuring and starting a new worker node. Of course in this time, added nodes have no data. HOG has to copy data on the new nodes or start to work without data locality. When the resource environment is getting more dynamic, the response time will be longer.

### 7.2 Hadoop on GPU

GPUs (graphic processing units) can be considered as a lot of parallel processors that their computation power and memory bandwidth is several times more than CPUs. General purpose computation on GPU (GPGPU) is a framework that users can develop correct and efficient GPU programs easily. In fact it is a technique to apply regular GPUs to execute general purpose applications instead of graphical processing. Recent developments in MapReduce programs are typically data – intensive and result sizes depend on data sizes. These two features present the following requirements for GPU programming:

- Appropriate thread parallelism to hide high latency and to utilize the memory bandwidth of the device.
- Pre – allocation of output buffers in the memory of the device for massive DMA transfer so that before GPU program starts, the GPU memory is allocated by the CPU.

As shown in figure 18, GPU can be modelled as a CPU with a lot of cores which contained some SIMD multiprocessor. In this model, system includes a host (a CPU) and one or more GPU devices. GPUs are considered as a massive amount of data – parallel co – worker processors.

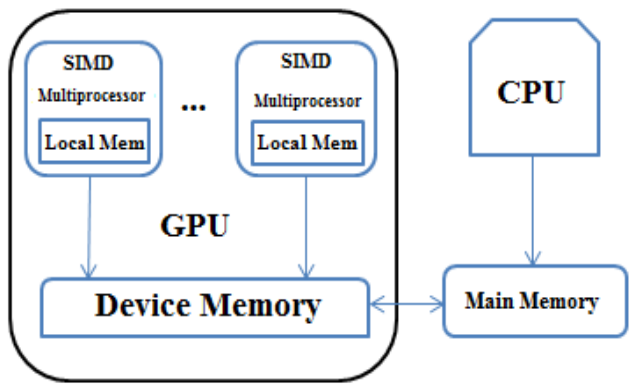

Fig 18: The Many-Core Architecture Model for GPUs.

Developing efficient GPU programs for complicated applications like applications with MapReduce is still a challenge, because GPUs have special purpose processor architecture and also each vendor has his own programming framework for complicated programs, but with integrated development environments like CUDA and OpenCL programmers can implement MapReduce tasks on GPUs. It is possible by Hadoop streaming which is an API that programmers can write their map and reduce functions in any language with standard I/O operations to implement their MapReduce program [83].

*7.2.1 MARS*

MARS is one of frameworks for accelerating MapReduce by GPUs. MARS is a MapReduce framework for simplify programming of complicated programs like programs that execute on MapReduce in GPU. In addition, MARS MapReduce framework makes possible combining codes which are accelerated by GPU with distributed environments like Hadoop and it does this by least effort [84].

In other word, MARS is merged with Hadoop and its result is MARSHadoop. In MARSHadoop each machine can use its CPU with Hadoop and also can use its GPU with MARSCUDA or MARSBROOK. It does not make any difference that MARS is running on which GPU or CPU, Cause application programming interface ( API ) for users is the same and it is like existing MapReduce systems based on CPU. Figure 19 depicts a workflow of GPU and CPU processing together.

In designing MARS, the MapReduce workflow is divided into 3 levels that are not very depending: group, map and reduce. Group level is designed for grouping map output through keyboard which is formatted for reduce input. Group level is like executing reduces with authentication function in original MapReduce system.

Since GPU does not support dynamic memory allocation during executing GPU codes, this limitation makes impossible to implement dynamic structures like queues and linked lists which are used in other MapReduce implementations. MARS provides a small set of APIs. Like existing MapReduce frameworks, MARS has two kinds of APIs: APIs which are implemented by user and APIs which are provided by system and users can use them as library calls.

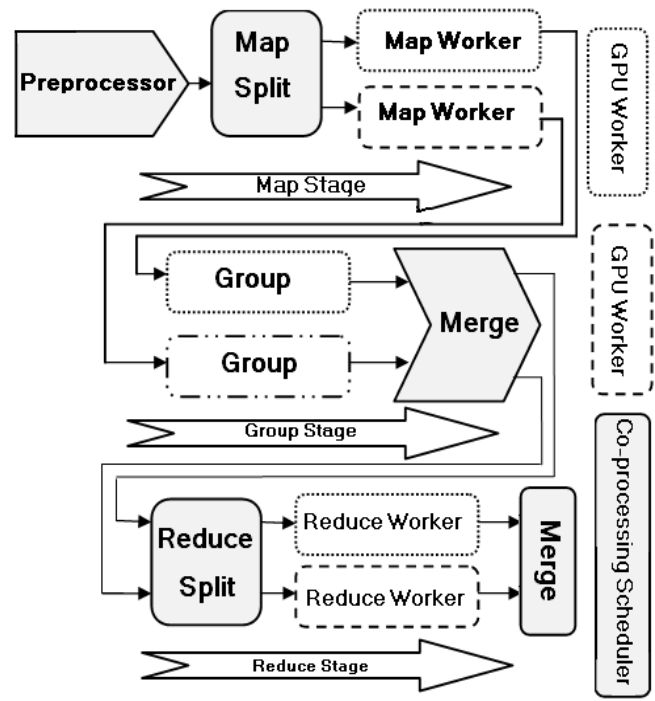

Fig 19: The Workflow of GPU/CPU Co – processing.

Hadoop uses CPUs on several machines and provides a distributed system with high – fault tolerance and other features. MARS uses GPUs to accelerate local computations. Hadoop streaming technology is used to combine MARS in Hadoop. This technology makes developers enable to use their own MapReduce implementations in Hadoop.

8. Security

What is security? Protecting information and property against stealing, corruption or natural disasters while allowing information and property stay available and efficient for some specified users. Hadoop has some levels of security. Current version of Hadoop has a very basic implementation of security. Actually it has a consulting access control mechanism. Hadoop does not really authenticate users. It authenticates users by executing "whoami" command in Linux. Everybody can communicate to DataNode directly without need to connect to NameNode and if somebody has the details of data block location, can ask for it. Hadoop clusters can be vulnerable against some kind of attacks like:[85]

- Unauthorized users can forge identity of authorized users and have access to cluster.
- Everybody can come cross the NameNode and get data block from DataNode directly.
- Eavesdropping/ sniffing of data packets which are sent to user by DataNode is possible.

But there are some ways to avoid these attacks:

- Authenticate the identity of users that have access to Hadoop by Kerberos protocol.
- Access certificate data on HDFS( granting and revoking capabilities)

In Hadoop massive amount of data are moving efficiency. Hadoop is not designed for real – time reading or updating. It carries data in high bandwidths. Databases in Hadoop answer the queries quickly but cannot match the bandwidth [86] .

The security in Hadoop must be optional, because all clusters are not shared between users. In the other hand it should not react to passwords quickly, cause this makes building Trojans easy. Hadoop must have only one sign in.

There are two definitions in security:

- Authentication: who is the user? In Hadoop version 0.2 the user is completely trusted. Users sent their username and group name over the wire. It is needed for both RPC and WEB UI.
- Authorization: what can that user do? HDFS has owner and group and permissions since 0.16. But MapReduce has nothing in 0.20.

Hadoop is extremely useful but it has security issues. For example Hadoop keep data in HDFS. The file system of version 0.19.3 has no control on reading and all jobs are executed as Hadoop user and file system does not follow access control list. As mentioned above, user identity in Hadoop can be forged. Hbase (BigTable for Hadoop) of version 0.19.3 has no critical access control measure. It does not control reading and writing. Every application can easily access data base by making a request for it. The problem is that access control is in the user level when it must be in the file system level. At the start of any read or write, access control list should be checked and user authentications must use a secure method like RSA key authentication [87].

Briefly, Hadoop has several security issues like:

- Insufficient authentication: Hadoop does not authenticate users or services properly.
- No privacy and lack of integrity: Hadoop uses insecure network transport and doesn't provide message level security.
- Arbitrary code execution: adversary users can submit jobs and execute them using the permissions of the TaskTracker [88].

## 8.1 A new approach in security

In 2009 security in Hadoop become an important topic. One goal of The Hadoop developers in 2010 was mutual authentication of users and services that would be transparent for users. In addition to changes in Hadoop core, Hadoop developers introduce a new workflow manager named Oozie [39]. They decided to use simple authentication and security layer (SASL) with Kerberos via GSSAPI to authenticate users for services. When a user contacts to JobTracker, that connection would be mutually authenticated by Kerberos. There are some flat configuration files that operation system principals in them are matched to a set of users and group access control lists [89].

In order to improve performance and prevent KDC to be bottleneck, developers use some tokens for communications which is secured by RPC digests. The new design of Hadoop security use agent tokens, job tokens and block access tokens. These tokens have same structure and are based on HMAC – SHA1. Users use agent tokens to communicate with NameNode to access to HDFS data. Access tokens are used for secure communications between NameNode and DataNode and also to make HDFS file system to export certificates. Job tokens are used to secure communication between MapReduce TaskTracker engine and tasks. This design use symmetric cryptography and shared key might be shared among hundreds or thousands of hosts [88].

## 8.2 Agent tokens

To prevent a flooding of authenticate requests at the beginning of a job, NameNode can build agent tokens. This allows users to authenticate only once and use the authenticating certificate for all jobs of a task. JobTracker automatically replicates tokens while a job is in process and cancels them when job is finished. Figure 20 shows the basic connection ways [86] .

In the time that new Kerberos mechanism and security were in use, Hadoop developers in Yahoo publish a new source code of workload named Oozie. Oozie allows users to simplify mission and management of MapReduce jobs. In order to enable Oozie to do this, a user Cloud would be choose and this user Cloud can do jobs from all users. Authentication is not implemented in Oozie. Of course there is a pluggable interface for Oozie, but there is no public authentication mechanism that is ready to use.

Anyone who wants to use Oozie must develop his own authentication mechanism. According to Hadoop security whitepaper [89], they considered writing an authentication plugin based on SPNEGO to support Kerberos authentication in browsers, but Jetty 6 limitations and supporting of various browsers prevent them to do this. In other presentations of Hadoop developers, the need for a default plugin authentication with priority for SPENGEO is discussed.

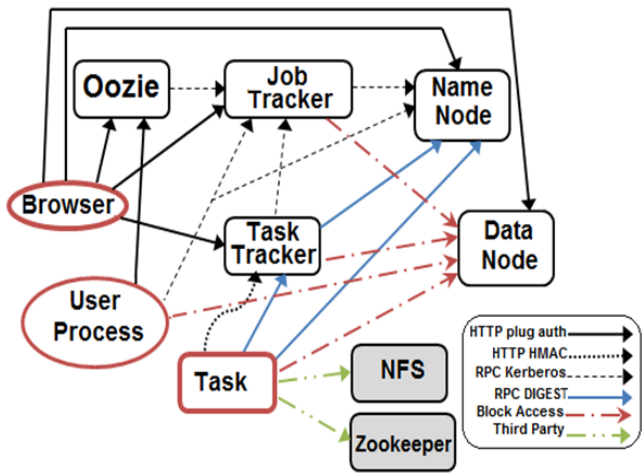

Fig 20: Basic Connection Ways of Agent Tokens in Hadoop.

To solve some of Hadoop objections, developers observe several issues. The new design needs end users don't have administrative rights on machines of cluster. If they have administrative access on machines in the cluster, they can access agent tokens, job tokens, block access tokens or symmetric cryptography key and destroy the guarantee of system security.

There are some concerns in developing security design for Hadoop; the most important of them is quality of protection (QoP). SASL is default. SASL framework which is added to support Kerberos in RPC communications can act much better. Other options for QoP protect the integrity and confidentiality of network communications. The default QoP for Hadoop is authentication which does not provide the integrity or confidentiality of network communications. This causes Hadoop RPC communications to be vulnerable against eavesdropping or manipulating.

The new design of Hadoop security is based on HMAC-SHA1 which is a symmetric key cryptography algorithm. In the case of block access tokens, symmetric key that is used in HMAC-SHA1 must be distributed between NameNode and DataNodes in the cluster which is potentially hundreds or thousands machines distributed geographically. If common key is revealed to an adversary, all data on all DataNodes would be vulnerable. With revealed data IDs, the adversary can have access to block access tokens and decrease Hadoop security to the previous level.

A lot of Hadoop services include HTTP. These services are: JobTracker, TaskTracker, NameNode, DataNode and new Oozie workload manager. To provide authentication for these web interfaces, Hadoop developers implement Web UI pluggable authentication. To do this, Hadoop end user must provide a Web UI authentication mechanism. There are multiple Web UIs such as Oozie, JobTracker and TaskTracker. But there is no standard HTTP authentication mechanism for them [88].

In some Hadoop deployments, HDFS proxies are used to transit massive data from server to server. Hadoop platform uses proxy IP addresses and a data base of lows to perform authentication and determine each user can do what. This can cause revealing of all data that HDFS proxy is authenticated to access them.

There are several suggested designs for Hadoop security. One of them is Hadoop on file system with least access (Tahoe) (Tahose – LAFS). Tahose – LAFS is a decentralised and open source storage which is trying to protect confidentiality and security, even if a specified server is in danger.

Hadoop on Tahose – LAFS assumes that disk is not trustable and network can't be trusted, but the memory of computational nodes can be trusted. Also unique nodes encrypt files on disks and communicate on a secure transition. Hadoop on Tahose – LAFS have an observable effect on Grid Mix performance, especially on writing.

*8.3 Web application proxy*

Web proxy application is part of yarn. It executed by default as a resource manager (RG) but can be executed alone. The reason of using proxy is to reduce the possibility of web based attacks to YARN.

In YARN, application master (AM) is responsible to providing a Web UI and send its link or RM. This have some potential issues: RM runs as a trusted user and persons who work for that web address and link to it, consider it trustable, while in reality AM runs as an un-trustable user and its link that is sent to RM can point to anything. Web proxy application reduce this risk and it does this by warn users to do not accept given application which is connected to an untrusted site.

In addition, Proxy also tries to reduce the effect that a destructive AM can have on a user. Basically this is done by scratching user cookies and replacing them with only one cookie which provides user name of the entered user. This is because most of the web based authentication systems recognize the user based on cookies. By giving this cookie to an unauthenticated application, potentially a door is open to abuse it. If cookie is designed correctly, this potential danger would be reduced fairly, but this is just reduction of that potential attack, current implementation of proxy does nothing to prevent running destructive JavaScript codes. In fact JavaScript can be used to get cookies. So currently, scratching cookies has little benefit.

9. Some Hadoop platforms and instances

Hadoop consists of several subs – projects or software layers. But basically it has two main parts: MapReduce and HDFS. There are some platforms which are built based on Hadoop platform. In this Section we introduce two of these platforms. Also we enumerate some of Hadoop usages.

## 9.1 Genie

Hadoop platform as an Amazon service: Amazon provides Hadoop infrastructure as a service via their elastic MapReduce (EMR). EMR provides an API to provision and run Hadoop clusters that users can run one or more Hadoop jobs on it. Genie provides a higher level of abstraction, where a client can submit individual Hadoop, Hive and Pig jobs via a rest – full API without having to provision new Hadoop clusters or installing Hadoop or Hive or pig clients. Also this enables administrators to manage the configuration of various back – end resources in the Cloud [90].

Genie is a set of rest – full services to manage jobs and resources in a Hadoop ecosystem. It has two key services: execution service which is a rest – full API for submitting and managing Hadoop, Hive or Pig jobs; And the configuration service which is a repository of available Hadoop resources with necessary Metadata to connect and run jobs on these resources.

Execution service: clients interact with Genie via the execution service API. They can launch a job by json or XML massages to this API and specify a set of parameters including:

- A job type which can be Hadoop, Hive or Pig
- Command – line arguments for the job
- File dependencies such as script and jar files ( e.g. for UDFS ) on S3
- A schedule type such as " ad – hoc " or " SLA " which Genie uses to map the job to an appropriate cluster
- A name for Hive Meta store to connect to (e.g. prod, test or one of the dev ones.)

If a job submission is successful, Genie returns a job id for the job, which can be used to get the job status, and the output URL. The output URL is an HTTP URL pointing to the working directory of the job, which contains the standard output and error logs. Each job id can translate to multiple MapReduce jobs depending on the number of intermediate stages in the Hive or Pig query being run.

The configuration service is used for tracking all running clusters and schedule types that they support.

Genie is used to manage resources dynamically.

## 9.2 Integrated Execution Platform (IEP)

Integrated Execution Platform (IEP) is an efficient consolidated platform for data streaming and Hadoop. Data streaming is processing incoming data from various resources like sensors, web click data, IP reports and so on, on the memory to realize real time computation. This computation model is getting more important since business organizations need to react to every event relevant to their avenue. Stream computation is a necessary computing model but it can only have access to data that resides in the memory. Many applications need to access to whole data to do a macro analysis. In addition, the concept of traditional batch computing like Hadoop only emphasizes high - throughput and does not consider latency strictly. IEP integrated system S for data streaming and Hadoop for batch processing to build a new platform. Its main goal is maximizing resource utilization. Developers of IEP extended hadoop to support a suspend/resume mechanism to control jobs and also they designed a dynamic load balancing for IEP to decrease failure detection time from a couple of minutes to a few seconds. They applied a time series prediction algorithm with relatively low computational complexity to load balancing. IEP increases CPU usage while keeping a low latency [91].

## 9.3 Some instances of Hadoop usages

As mentioned before, Hadoop is most popular platform for storing and processing BigData. It is used in a lot of scientific areas. In this Section we introduce a few examples of Hadoop usages.

### 9.3.1 Molecular dynamic simulation

Hadoop is an open – source distributed computing environment with simplified designing and developing parallel programming and allows developers to focused on analysing complex applications and do not bother with details of parallel implementation. For these reasons Hadoop has been widely used in bioinformatics. Traditionally, molecular dynamic usually runs on MPI (message passing interface) and has big issues like lack of fault tolerance and load balancing problems. Executing molecular dynamic simulator on Hadoop has some advantages like solving traditional software and hardware storage problems, short simulation time and using Hadoop capabilities in scientific computing. One instance of using Hadoop in bioinformatics is using it in next – generation sequencing. The Cloudburst software uses Hadoop for analysis of next – generation sequencing data and then maps the next generation short sequencing data into a reference genome for SPN discovery and genotyping [43], [56].

### 9.3.2 Log files and click stream analysis

Log processing is one of most regular problems that are solved by Hadoop. Large data centers can generate several gigabytes of logs every day. These logs may be useful for solving problems and analysis, but storing and processing such amount of data is a serious technical challenge. Even a simple search when it is done in such massive data, will turn to a complex issue. A single machine can do this kind of work efficiently. This shows the necessity of using clusters of distributed machines to work on these tasks. This is where Hadoop is useful since it provides tools for storing (HDFS) and processing (MapReduce) large amount of log files [58].

### 9.3.3 Suggestion engines with commercial purposes

In order to suggesting most appropriate commercial for a special user, all existed and available information about that user is needed to be processed (like profile, web search history, click history, email headers and so on). Suggestion engine should be able to understand behaviour and performances of each user and be able to estimate probability of interests of user to each commercial. Another kind of analysis is that an organization process all available data of resources related to each user to understand what to do to increase the user satisfaction ( internal travels, useful programs, personal behaviours and so on). Such works can be done efficiently by Hadoop, since Hadoop allows different part of data (data about each user) be analysed separately and in parallel.

*9.3.4    Generating search index*

This is where Hadoop was built originally as an effort to build an open – source engine. After Google published its GFS [77] and MapReduce papers, Hadoop was designed and implemented. The problem was sorting and processing a massive amount of related data. The main result of this process is building search index. Every pieces of data ( like web sites) can be processed separately from other portions of data, but in the heart of programming and industry, there is a need to processing and storing massive amount of data and it is where HDFS / MapReduce becomes very useful.

In GFS there is only one master GFS node (to avoid the cost of coordination between master nodes) and several client or chunkserver nodes. Each chunkserver communicate with master server only for Metadata and chunkservers communicate with each other for other data [77]. Researches have shown that indexes can improve the performance of MapReduce; however the existing approaches for index creations have high costs. A new approach in [92] designed a novel indexing in Hadoop named HAIL (Hadoop Aggressive Indexing Library)[63] which creates different cluster of indexes over massive data with very low costs and almost zero overhead for indexing. It improves the runtimes of several classes of MapReduce jobs.

*9.3.5    Optimizing Sequence Alignment in Cloud Using Hadoop and MPP Database*

Hadoop is used in bioinformatics. In this area of science, sequence alignment is the method of arranging the sequences of DNA or RNA or a part of them to find similarities between them that might declare a relationship between sequences. This kind of information is used in medical and biological researches only if this information can be analysed.  In order of obtaining functional revelation the factors that must be considered while aligning sequences are: high – speed matching, optimized querying of sequences and accuracy of alignments. Since data growth every day and processing massive data required new technologies, in a new approach, Greenplum Massively Parallel Processing (MPP) database and Hadoop are combined to perform parallel technology for optimizing sequence alignment. The Figure 21 shows an example of reading data from HDFS into the Greenplum database.[93]

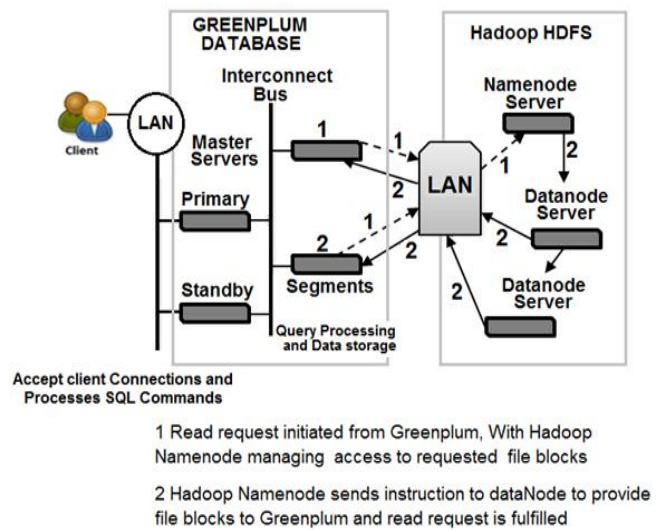

Fig 21: Example of read from HDFS into Greenplum Database.

*9.3.6    Hadoop in Mobile Computing*

There is a framework for mobile computing based on Hadoop which is named Hyrax. In Hyrax there is a central server (NameNode) that has access to other nodes (DataNodes) and coordinates data and jobs. Phones communicate to each other through an 802.11g network. Like general Hadoop, NameNode does not do any processing. Each client phone (DataNode) executes instances of DataNode and TaskTracker and also runs threads that store phones data on HDFS and threads for recording sensor data. In Hyrax if node failure is anticipated, its task will be replicated to other nodes and makes the whole system stable[7].

There are also another frameworks based on Hadoop for mobile computing. Some of them uses Hadoop MapReduce framework for receiving and processing sensor data in a virtual mobile Clouds and use these information to distinguish the location of the device (to use them in place bound tasks).  This information can be used to manage sending and receiving jobs to and from other devices [7].

In mobile computing, disconnectivity is a normal problem which is why Hadoop is useful in this area: Hadoop is a fault tolerant platform and hardware failures are common in HDFS architecture.

## 10. Hadoop problems and challenges

Despite the fact that in a large- scale distributed system such as Hadoop, if some nodes misbehave, other nodes and entire system may still work properly, but debugging whole system is a challenge; Because many performance problems are unpredictable and transient. Hadoop updates quickly and nearly every month, comes up with a new version and in each new version, some problems are solved. But there are no major changes in its network protocols. Hadoop is a parallel system and it is not suitable to diagnose system by processing time. The normal interval between two messages in Hadoop can be varying quickly, from seconds to minutes. So it is difficult to find a single diagnosis algorithm for all applications [94]. There are some major problems of hadoop that we describe some of them in following paragraphs.

1- In general deployment of Hadoop, there is only one NameNode/JobTracker: Hadoop uses a special master server to run tasks on the sub servers. This causes some problems like fatal errors that result in system failure and also memory capacity shortage which endanger scalability seriously [1], [41].
2- Small HDFS files: HDFS data are stored as objects and each object takes about 150 Byte space. If we have a lot of these small files to store, NameNode would need a lot of memory. This problem restricts Hadoop scalability.
3- JobTracker may have to bear too much load, because it has to monitor and distribute simultaneously. Yahoo starts to build a new generation of Hadoop MapReduce to solve this problem. The goal of Yahoo efforts is to separate monitoring from distribution by building a special component for monitoring while JobTracker is just in charge of scheduling tasks.
4- Data processing performance: Hadoop is like a data base and need special optimizations based on application needs and demands. Many experiments show that there is still a lot of space for improvement.
5- Hadoop scheduler doesn't exploit data locality or partitioning skew present in some applications while schedules reduce tasks. This might increase the cluster network traffic. There are some studies to solve this problem by designing new scheduler for schedule reduces tasks [95].
6- Hadoop does not work efficiently in typical wide – area network topologies which PCs in them, are hidden behind firewall and NAT [18].
7- In MapReduce design, all keys generated by Map phase must fit into the memory, but there are some workloads that have a distribution of keys which irritate the growth of intermediate data structure and so may does not fit in available main memory. This may cause the page fault caused by the use of swap area [16].
8- Hadoop is limited in SQL support, because it lacks some basic SQL functions.
9- Hadoop does not have an efficient plan for execution and it results in that the clusters in Hadoop are usually larger than clusters in similar databases [59].
10- Hadoop MapReduce does not have a flexible resource management and cannot react quickly to user or application demands. Also managing multiple application integrations on production – scale distributed system is difficult. There is no automated application service deployment capability in it [96].

Also from other perspective, the errors that occur in Hadoop can be classified as bellow:

- Operational error: missing or incorrect operations and artifacts (files and directories); this includes errors during restarting or shutting down of a node, files not created or even created with wrong permission or moved or operations that cause inconsistency of system.
- Configuration errors: errors like syntax errors or illegal and lexical errors in software systems. These errors might cause inconsistency in ecosystem [60].
- Software errors: compatibly issues among various part of the system (like Hbase and HDFS).
- Resource errors: resource unavailability which cause system failures [55].

For users even operational professionals who have a little experience with Hadoop, the deployment and using of Hadoop can be very error – prone and diagnosis of causes can take a lot of time. There are a lot of things that must be done correctly to set up a Hadoop cluster and in the end, for beginner, it might still does not work. For example, if someone wants to use Hadoop, he should learn Java an MPP and have knowledge about MapReduce and so on [59].

Apache Chukwa (Hadoop log aggregation) project is a monitor for measuring the performance of Hadoop. It is a cluster monitoring system which uses HDFS and MapReduce for processing and storage. Chukwa is a general purpose monitor and it can monitor a Hadoop cluster perfectly, but it cannot concentrate on analysis of individual jobs [38],[39].

There is a lack of MapReduce optimizer in Hadoop. The jobs in MapReduce are scan – oriented, because the whole platform is designed to executing long running jobs. Also some classes of tasks do not fit naturally in MapReduce paradigm. For example, updates, joint and iterative tasks cannot be well expressed in MapReduce. Furthermore, the performance of Hadoop MapReduce in some cases is far behind of optimized parallel DBMS [63].

Some researchers have shown that Hadoop works well in homogeneous environments, but it does not work well in heterogeneous Cloud computing infrastructures. They showed that the assumption of homogeneity in some cases (like Amazon EC2) can lead to wrong and unnecessary speculative execution when environment is not homogeneous. This assumption may result in worst performance compare to when speculation id disabled. In order to solve this problem, Cloud execution management systems should be designed to deal with heterogeneity [23].

While Hadoop is great in some distributed data processing, it is not a suitable platform for some areas like highly interactive analytic process. Hadoop makes multiple copies of already BigData and so decreases efficiency, especially in analytic processing. It also has limitations in Database functionality. These kinds of gaps in Hadoop are because of its open source distributed nature. In open source developments, quality assurance is not a big concern of developers. Also HDFS is a file system which is designed to operate on arbitrary sized clusters, so transaction consistency or recovery checkpoints are not discussed in it. It means that for some jobs, the result of the process may not be 100 percent correct [59].

Hadoop sub - projects have an important effect on its performance; for example, user – level file systems that are mounted on a client which uses Fuse, bear a lot of I/O overhead due to extra memory copies and context switches during local file access. Because Fuse forces applications to access data in the mounted file system through the userland daemon, even when data is stored locally. Also redundant data are stored in multiple page caches and increase memory usages which are a big concern for storing already BigData. These overheads are big and can degrade the performance of data – intensive applications [36].

### 10.1 Some challenges for MapReduce in Clouds

There are some challenges for MapReduce in Clouds to control the performance. Clouds usually provide a variety of options for storage, like off – instance Cloud storage, mountable off – instance Cloud storage or virtualized Cloud storage which continue for lifetime of the instance and can be used to set up a file system like HDFS. Choosing the best MapReduce deployment is very important, because the performance of data – intensive applications depends on the storage location and bandwidth [64].

MapReduce framework needs to keep Metadata information to manage the jobs as well as the infrastructures. These Metadata should be stored reliably and provide good scalability and accessibility to prevent the single point of failures and bottlenecks for MapReduce processes. The HDFS center stored the index information which is called Metadata and uses this Metadata to manage several partitions of Big Data which are distributed to a lot of nodes in a cluster [62].

Cloud infrastructure has inter – node I/O performance fluctuations because networks are shared and topology is not defined and this affect the intermediate data transfer performance of MapReduce applications, in other word, this can influence the communication consistency and scalability.

Clouds are implemented as shared infrastructure operating that use virtual machines, so their performance is affected by the load of underlying infrastructure services as well as the load of user applications sharing nodes which hosts the virtual machines.

Cloud shave a good reliability, however node failures are common when a large number of nodes are used for computation, these node failures are more common when virtual instances are running on top of non – dedicated hardware. MapReduce framework can recover jobs from worker nodes failure easily because of the replicate factor, but this problem can become a fatal failure when it happens in the master node, cause master node stores Metadata and is in charge of job scheduling queues and so on.

Cloud instance storage is maintained during the lifetime of the instance and after that, the information logged to the instance storage would be deleted, so no one can process the log afterward if this information is needed. However performing excessive logging to a limited bandwidth can become a performance bottleneck for MapReduce computation.

### 10.2 Optimization in Cloud

In every area which Hadoop is used, there are different kinds of challenges that might be considered to gain the best optimality and utilization. Hadoop has more than 180 configuration parameters like the number of replicated input data, the number of parallel MapReduce tasks, the number of parallel connections for data transition and etc., so the number of MapReduce tasks in each resource set must be carefully choose to maximize utility and optimize the performance [97].

Every application has a different resource bottleneck and different resource utilization, and needs a different combination of number of map and reduces tasks to gain the maximum resource utilization. When there are large values of chosen parameters, this may result in resource contention and lower overall performance; on the other hand statically choose small values may cause the underutilization of resources.

There are some approaches to utilize Hadoop on Cloud. The Figure 22 shows one of these methods. The figure consisted of two components: the first is RS (resource set) maximize which is responsible for calculating the optimum parameters for Hadoop jobs to make each resource utilized and the second is RS Seizer which determines the number of resource sets which is needed to minimize the cost while the performance is maximized.

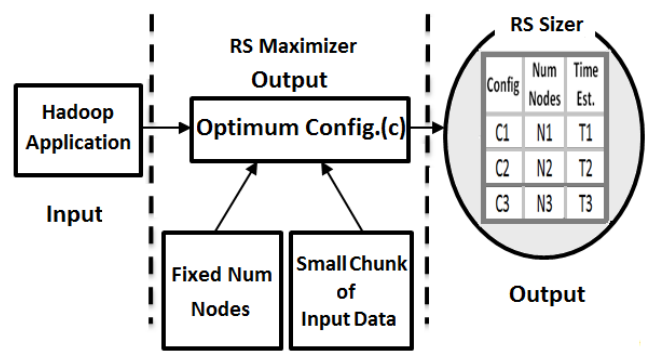

Fig 22: An Approach of Hadoop Optimization in Cloud.

## 11. Conclusion

In this paper we surveyed a technology for massive data storage and computing named Hadoop. Hadoop becomes most popular system for processing large data sets and its performance is very impressive. Hadoop consists of heterogeneous computing devices like regular PCs or other commodity resources. A Hadoop cluster is made of two parts: HDFS and MapReduce. It uses HDFS for data management. HDFS provides storage for input and output data in MapReduce jobs and is designed with abilities like high - fault tolerance, high - distribution capacity and high - throughput. It is also suitable for storing Terabytes or Petabytes data on cluster and it runs on flexible hardware like PCs. Hadoop is a technology that can be developed by high - efficiency, high - speed and low - cost systems and meets today's demands for resources and services. We described different aspects of Hadoop MapReduce platform including architecture, schedulers and security and some of its problems and challenges. Also we illustrated some of Hadoop platforms and some instances of Hadoop in real world.

While Hadoop made a revolution in storage and processing BigData, it still has some challenges. Some of these challenges are due to its open source distributed nature. Hadoop is a set of projects and each project is created to solve a problem, so quality assurance in all of them is very difficult and its results are usually with a solution of unpredictable usability and quality. Also Hadoop doesn't work well in some areas like highly interactive analytics.

Every day, some of Hadoop problems are solved. Since it consists of several subs – projects, optimization in each of them makes Hadoop more efficient.